\documentclass[twocolumn]{aa} 

\usepackage{graphicx}
\usepackage[varg]{txfonts}
\usepackage[percent]{overpic}
\usepackage{color}
\usepackage{multirow}
\usepackage{natbib,twoopt} 
\usepackage[colorlinks=true,urlcolor=blue,linkcolor=red]{hyperref} 
\bibpunct{(}{)}{;}{a}{}{,} 
\begin{document}   
   \title{Absolute velocity measurements in sunspot umbrae}
   
   \author{J. L\"ohner-B\"ottcher\inst{\ref{inst_kis}} \and W. Schmidt\inst{\ref{inst_kis}} \and R. Schlichenmaier\inst{\ref{inst_kis}} \and H.-P. Doerr\inst{\ref{inst_mps}} \and T. Steinmetz\inst{\ref{inst_mpq},\ref{inst_menlo}} \and R. Holzwarth\inst{\ref{inst_mpq},\ref{inst_menlo}}}
    \institute{
    	Kiepenheuer-Institut f\"ur Sonnenphysik, Sch\"oneckstr. 6, 79104 Freiburg, Germany\\ \email{jlb@leibniz-kis.de}\label{inst_kis} \and
	Max-Planck-Institut f\"ur Sonnensystemforschung, Justus-von-Liebig-Weg 3, 37077 G\"ottingen, Germany\label{inst_mps} \and 
	Max-Planck-Institut f\"ur Quantenoptik, Hans-Kopfermann-Strasse 1, 85748 Garching, Germany\label{inst_mpq} \and
	Menlo Systems GmbH, Am Klopferspitz 19, 82152 Martinsried, Germany\label{inst_menlo}}
   \date{Received 23 Feb 2018 / Accepted xx xxx 2018}

  \abstract 
  {In sunspot umbrae, convection is largely suppressed by the strong magnetic field. Previous measurements reported on negligible convective flows in umbral cores. Based on this, numerous studies have taken the umbra as zero reference to calculate Doppler velocities of the ambient active region.}
  {To clarify the amount of convective motion in the darkest part of umbrae, we directly measured Doppler velocities with an unprecedented accuracy and precision.} 
  {We performed spectroscopic observations of sunspot umbrae with the Laser Absolute Reference Spectrograph (LARS) at the German Vacuum Tower Telescope. A laser frequency comb {enabled the calibration of} the high-resolution spectrograph and absolute wavelength positions for 13 observation sequences. A thorough spectral calibration, including the measurement of the reference wavelength, yielded Doppler shifts of the spectral line \ion{Ti}{i}\,5713.9\,\AA\ with an uncertainty of around $5\,{\rm m\,s^{-1}}$. A bisector analysis gave the depth-dependent line asymmetry.}
  {The measured Doppler shifts are a composition of umbral convection and magneto-acoustic waves. For the analysis of convective shifts, we temporally average each sequence to reduce the superimposed wave signal. Compared to convective blueshifts of up to $\mathrm{-350\,m\,s^{-1}}$ in the quiet Sun, sunspot umbrae yield a strongly reduced convective blueshifts around $\mathrm{-30\,m\,s^{-1}}$. {W}e find that the velocity in a sunspot umbra correlates significantly with the magnetic field strength, but also with the umbral temperature defining the depth of the \ion{Ti}{i}\,5713.9\,\AA\ line. The vertical upward motion decreases with increasing field strength. Extrapolating the linear approximation to zero magnetic field reproduces the measured quiet Sun blueshift. In the same manner, we find that the convective blueshift decreases as a function of increasing line depth.}
  {Simply taking the sunspot umbra as a zero velocity reference for the calculation of photospheric Dopplergrams can imply a {systematic} velocity error reaching $\mathrm{100\,m\,s^{-1}}$, or more. Setting up a relationship between vertical velocities and magnetic field strength provides a remedy for solar spectropolarimetry. We propose a novel approach of substantially increasing the accuracy of the Doppler velocities of a sunspot region by including the magnetic field information to define the umbral reference velocity.}

  \maketitle
  \titlerunning{Absolute velocity measurements in sunspot umbrae} 
  \authorrunning{L\"ohner-B\"ottcher et al.}

\section{Introduction}\label{sec1_intro}
In the solar interior, the energy transport toward the solar surface occurs via convection. When measured spectroscopically in the unresolved quiet Sun, overturning convective motions lead to an overall blueshift of photospheric spectral lines. Upflows in the bright granular cells have a larger contribution to the composed line profile than downflows in the dark intergranular lanes. The line profile then shows an asymmetric, often "C"-shaped bisector which is associated with height-dependent convective motion. In a sunspot umbra, the magnetic field is strong enough to suppress this overturning convection. Since the convective heat flux is strongly reduced, the umbra appears substantially darker and significantly less blueshifed than its vicinity. From an observational and theoretical point of view, it still remains controversial whether convection is suppressed entirely in the darkest part of the umbra. In the absence of convection, the convective shift would then become zero and the spectral line shape should become symmetrical. However, to account for the observed brightness of a sunspot umbra some residual energy transport is necessary, at least at deep atmospheric layers. If unresolved convective motions would provide the energy, \citet{1989ApJ...342.1158M} expect a convective blueshift of around $\mathrm{-150\,m\,s^{-1}}$ in the deep umbra. An absence of significant blueshifts would imply that most energy must be transported by radiation \citep{1991ApJ...373..683L}, although some convective motions (probably in umbral dots) must exist since radiation alone cannot explain the umbral heat flux \citep{1965ApJ...141..548D}.

Direct observations of absolute velocities in the umbra have always been technically difficult. The measurement of convective Doppler shifts in umbrae require spectroscopic observations with a high spectral resolution, the selection of a suitable spectral line with a well-known laboratory wavelength, a careful reduction of any superimposing systematic Doppler shift, and an accurate wavelength reference. In addition, the telescope and the atmospheric conditions have to guarantee a high spatial resolution in order to decrease the amount of scattered light. Additionally, observational time series have to be long enough to filter umbral 5\,min oscillations. 

\citet{Beckers1977} was the first to measure absolute velocities in sunspot umbrae. He performed his observation with the 40\,cm coronagraph and the large Littrow spectrograph at the Sacramento Peak Observatory. The spectral \ion{Ti}{I} line at 5713.9\,\AA\ was selected since it becomes stronger in cooler atmosphere and features no Zeeman splitting due to the Land\'e factor of $\mathrm{g_{eff}=0}$. Using an iodine vapor absorption tube, developed at the Sacramento Peak Observatory in the mid 70s \citep{1980A&A....81...50B}, the absolute wavelength calibration reached an accuracy of $\mathrm{\pm30\,m\,s^{-1}}$. \citet{Beckers1977} observed six sunspots at various heliocentric positions while crossing the solar disk. Besides the fact that the observed umbrae yielded no center-to-limb variation of the Doppler shift, the vertical flow inside the umbrae appeared to be small. Ignoring a possible (small) pressure shift of the spectral line, the blueshift had an upper limit of $\mathrm{-25\,m\,s^{-1}}$. In the absence of the typical signatures of convective motion, sunspot umbrae were therefore assumed to be at rest. 

\citet{1984SoPh...93...53K} performed a quite similar sunspot study also using an iodine cell spectrum for the absolute wavelength calibration. The measured blueshifts of the \ion{Fe}{I}\,5576\,\AA\ line ($\mathrm{g_{eff}=0}$) reached up to $\mathrm{-140\,m\,s^{-1}}$. Essentially supporting the thesis of umbrae being at rest, the author argued that a large amount of stray light from the surrounding quiet Sun could have affected the Doppler shift leading to a stronger blueshift and more asymmetric line profile. In addition to that, the laboratory wavelength of the \ion{Fe}{I} line was not known with a sufficient accuracy. 

In the last decades, many studies on umbral and penumbral velocities made use of telluric lines in the proximity of solar lines (commonly at 6302\,\AA) for their absolute wavelength calibration \citep[e.g.,][]{1991ApJ...373..683L,Schmidt+Balthasar1994,1997ApJ...474..810M,1999A&A...349..941S,2006A&A...454..975R,2008ApJ...676..698B}. Under the assumption of telluric lines being invariant and a valid determination of the solar spectral line shift \citep{Balthasar+etal1982}, most studies found either negligible or moderate blueshifts of several tens to one hundred $\mathrm{m\,s^{-1}}$ for the central umbra. The reported uncertainties of the Doppler velocities were typically around $\mathrm{\pm150\,m\,s^{-1}}$. It was concluded that the inner umbra is at rest except for umbral oscillations.

Various indirect measurements of umbral (and penumbral) velocities have been performed by taking an inherently solar reference. One way of calibrating the Doppler velocities of a sunspot region has been by taking the quiet Sun and its convective blueshift as the reference \citep[e.g.,][]{1994ApJ...430..413S,2004A&A...427..319B,2004A&A...415..717T,2009A&A...508.1453F}. The {spectra from the} granular vicinity of the sunspot {are} averaged, and the center of the line profile is attributed to the nominal convective blueshift. By calculating the relative Doppler shift of the umbral profile, most studies yield a blueshift between $\mathrm{-50\,m\,s^{-1}}$ and $\mathrm{-120\,m\,s^{-1}}$ for the umbral core. The uncertainty of at least $\mathrm{\pm100\,m\,s^{-1}}$ arises from two factors. Firstly, the reference values of the convective blueshift stems either from observations \citep{1981A&A....96..345D,1984SoPh...93..219B} or numerical models \citep{2002A&A...385.1056B,2007ApJ...655..615L,2011A&A...528A.113D} and have an estimated error of around $\mathrm{\pm50\,m\,s^{-1}}$. Secondly, the direct surroundings which have to be assumed as quiet Sun may be affected by magnetic fields or flow fields (moat, supergranulation) which cause additional errors.

In case of spectropolarimetric sunspot measurements, another indirect way to obtain Doppler velocities is to define the zero-crossing wavelength of the Stokes V profile of a magnetically sensitive line as the zero velocity reference. Commonly, the average Stokes V profile of the dark umbra is taken as the zero-velocity reference \citep{2009A&A...508.1453F,2015ApJ...803...93E}, with uncertainties considered to be around $\mathrm{\pm100\,m\,s^{-1}}$. On the other hand, the Stoke V zero-crossing point from the plage region around the sunspot has been determined to be very close to the laboratory wavelengths of the spectral line \citep{1986A&A...168..311S,1997ApJ...474..810M}. Following this approach, \citet{2005ApJ...622.1292S} detected weak upward motions around $\mathrm{-70\,m\,s^{-1}}$ at the inner umbra. However, the systematic error was estimated to $\mathrm{\pm250\,m\,s^{-1}}$.

Qualitative studies of penumbral Evershed flows or small-scale dynamics in umbral dots require the determination of accurate Doppler velocities. Based on the findings of \citet{Beckers1977}, numerous studies on sunspot dynamics have taken the darkest, most homogenous part of the umbra as their zero velocity reference \citep[e.g.,][]{1994A&A...290..972R,1997ApJ...477..485S,2013ApJ...770...74K}. While dark umbrae may provide a valid zero velocity reference, some studies applied the zero velocity assumption even to the areal average of pores with much weaker magnetic fields \citep{2010ApJ...713.1282O}. 

In this work, we aim to verify whether sunspot umbrae are actually at rest. To this end, we performed high-resolution spectroscopy with LARS at an unprecedented spectral quality. The sunspot observations and spectroscopic data are described in Section \ref{sec2_data}. The spectral investigation of the \ion{Ti}{I}\,5713.9\,\AA\ allows a direct comparison with the results of \citet{Beckers1977}. Enabled by adaptive optics and a laser frequency comb {calibrating} the high-resolution echelle spectrograph \citep{Steinmetz+etal2008,Doerr2015}, our measurement accuracy outperforms previous studies by about an order of magnitude. In Section \ref{sec3_results}, we present the umbral Doppler velocities and discuss their dependence on other parameters, like the spectral line depth and the magnetic field strength. In our conclusions in Section \ref{sec4_conclusions}, we propose a careful usage of sunspot umbrae as the velocity reference for the calibration of sunspot Dopplergrams.

\begin{table*}[htbp]
\caption{List of sunspot observations and resultant Doppler velocities of the umbral cores.}
\label{table_observations}
\centering
\begin{tabular}{cccccccccccccc}
\hline\hline
Obs.&Sunspot&Date&\multicolumn{2}{c}{Position}&\multicolumn{2}{c}{Heliocentric}&$\rm B_{abs}$&$\rm t_{cycle}$&Cycles&Time&${\rm v_{los, \pm50\,m\AA}}$&${\rm v_{los, \pm25\,m\AA}}$\\ 
\#&(NOAA)&&X&Y&$\alpha$&$\mu$&(kG)&(s)&&(min)&(${\rm m\,s^{-1}}$)&(${\rm m\,s^{-1}}$)\\
\hline
1&12150&2014\,/\,08\,/\,29&168\arcsec&$-$333\arcsec&$23.1^{\circ}$&0.92&2.32&2.5&647&27&$-17.3$&$-40.9\,\mathbf{(\pm6.5)}$\\
2&&2014\,/\,08\,/\,26&$-$452\arcsec&$-$316\arcsec&$35.5^{\circ}$&0.81&2.32&2.5&800&33&$-25.0$&$-49.7\,\mathbf{(\pm5.5)}$\\
3&&2014\,/\,08\,/\,25&$-$608\arcsec&$-$305\arcsec&$45.8^{\circ}$&0.70&2.25&2.5&800&33&$-42.4$&$-61.8\,\mathbf{(\pm5.3)}$\\
&&&&&&&\multicolumn{3}{r}{Total (12150)}&94&$-28.9$&$-51.5\,\mathbf{(\pm5.2)}$\\
\hline
4&12149&2014\,/\,08\,/\,28&267\arcsec&51\arcsec&$16.6^{\circ}$&0.96&2.75&2.5&800&33&$+16.0$&$-18.0\,\mathbf{(\pm5.6)}$\\
5&&2014\,/\,08\,/\,25&$-$385\arcsec&64\arcsec&$24.3^{\circ}$&0.91&2.50&2.5&800&33&$-26.4$&$-52.3\,\mathbf{(\pm5.5)}$\\
&&&&&&&\multicolumn{3}{r}{Total (12149)}&67&$-5.2$&$-35.2\,\mathbf{(\pm5.3)}$\\
\hline
6&12146&2014\,/\,08\,/\,26&715\arcsec&61\arcsec&$49.1^{\circ}$&0.65&2.43&2.5&780&32&$-2.6$&$-27.4\,\mathbf{(\pm5.4)}$\\
7&&2014\,/\,08\,/\,25&548\arcsec&46\arcsec&$35.4^{\circ}$&0.81&2.56&2.5&800&33&$+9.6$&$-15.7\,\mathbf{(\pm5.6)}$\\
8&&2014\,/\,08\,/\,24&361\arcsec&36\arcsec&$22.5^{\circ}$&0.92&2.47&2.5&800&33&$+1.3$&$-21.5\,\mathbf{(\pm5.8)}$\\
9&&2014\,/\,08\,/\,24&351\arcsec&36\arcsec&$21.8^{\circ}$&0.93&2.59&2.5&800&33&$+9.8$&$-16.3\,\mathbf{(\pm5.5)}$\\
10&&2014\,/\,08\,/\,23&204\arcsec&30\arcsec&$12.6^{\circ}$&0.98&2.46&2.5&800&33&$+3.7$&$-28.4\,\mathbf{(\pm5.5)}$\\
11&&2014\,/\,08\,/\,21&$-$287\arcsec&40\arcsec&$17.8^{\circ}$&0.95&2.50&2.5&1426&59&$-8.5$&$-27.8\,\mathbf{(\pm5.4)}$\\
&&&&&&&\multicolumn{3}{r}{Total (12146)}&225&$+1.0$&$-23.4\,\mathbf{(\pm5.1)}$\\
\hline
&&&&&&&\multicolumn{3}{r}{Total (above)}&415&$-5.2$&$-29.8\,\mathbf{(\pm5.0)}$\\
\hline
12&12109&2014\,/\,07\,/\,13&855\arcsec&$-$161\arcsec&$67.2^{\circ}$&0.39&2.72&2&900&30&$+97.2$&$+51.3\,\mathbf{(\pm5.1)}$\\
13&&2014\,/\,07\,/\,12&770\arcsec&$-$171\arcsec&$56.7^{\circ}$&0.55&2.86&2&900&30&$+74.2$&$+37.0\,\mathbf{(\pm5.1)}$\\
&&&&&&&\multicolumn{3}{r}{Total (12109)}&60&$+85.7$&$+44.2\,\mathbf{(\pm5.0)}$\\
\hline
&&&&&&&\multicolumn{3}{r}{Total (all)}&475&$+6.3$&$-20.5\,\mathbf{(\pm5.0)}$\\
\hline
\end{tabular}
\tablefoot{The table lists the active region number, the observation date, the heliographic position (X,\,Y) on the solar disk, the heliocentric angle $\alpha$ and parameter $\mu=\cos\alpha$, and the absolute magnetic field strength $\rm B_{abs}$. The observed spectroscopic sequences are defined by the cycle time $\mathrm{t_{cycle}}$ and the total number of cycles which amount to the total time of the sequence. The temporally averaged Doppler velocities ${\rm v_{los}}$  base on fit ranges of $\mathrm{\pm50\,m\AA}$ and $\mathrm{\pm25\,m\AA}$ around the \ion{Ti}{i}\,5713.9\,\AA\ line center. The total uncertainty (in brackets) includes systematic and statistic errors.}
\end{table*}

\section{Observations with LARS}\label{sec2_data}
During two observation campaigns in 2014 at the Observatorio del Teide on Tenerife, we performed sunspot observations with the Laser Absolute Reference Spectrograph \citep[{LARS},][]{Doerr2015,2017A&A...607A..12L} at the German Vacuum Tower Telescope \citep[{VTT},][]{1985VA.....28..519S}. In the following, we give an overview of the observed sunspots and a detailed description on the spectroscopic measurements with LARS.
 
\subsection{Sample of sunspots}\label{sec2_spots}
Between July 12 and August 29 2014 we observed four different sunspots (in active regions NOAA\,12109, 12146, 12149, and 12150). As listed in Table\,\ref{table_observations}, each sunspot was observed at least twice. In total, we have performed 13 individual observation sequences. The sunspots were located at heliocentric angles $\alpha$ preferably smaller than $50^{\circ}$, except for NOAA\,12109. The sunspots were stable and had a fully or almost fully developed penumbrae. All umbrae had a diameter of at least 10\arcsec\ and an absolute magnetic field strength exceeding 2.2\,kG. The latter information was extracted from the magnetic field inversions \citep{2011SoPh..273..267B} from the Helioseismic and Magnetic Imager \citep[HMI,][]{2012SoPh..275..229S} instrument on the Solar Dynamics Observatory \citep{2012SoPh..275....3P}.

\subsection{Spatial context information}

\begin{figure}[htpb]

\includegraphics[trim=1.7cm 4.9cm 9.75cm 3.25cm,clip,width=\columnwidth]{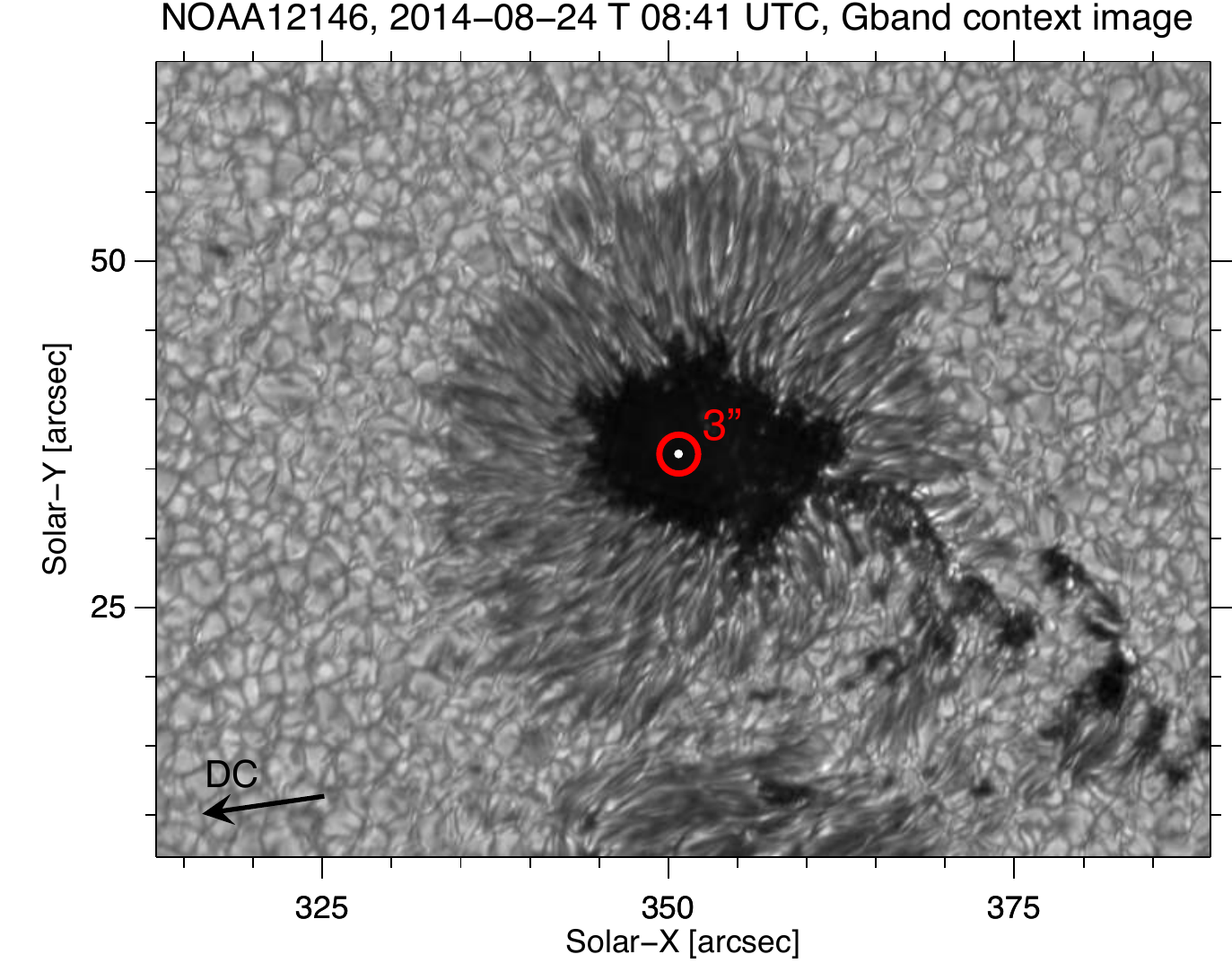}
\caption{Sunspot of NOAA\,12146 observed with the LARS context imager on August 24 2014 at 8:41\,UTC in the spectral G-band at 4307\,\AA. The black arrow is pointing in the direction of the solar disk center. The spectroscopically analyzed 3\arcsec-wide region (highlighted by the red circle) was centered to the darkest part of the umbra at $\mathrm{X=361\arcsec}$ and $\mathrm{Y=36\arcsec}$. For more information see Obs.\,8 in Table\,\ref{table_observations}.}
\label{fig_results_Ti5714_spot}
\end{figure}

Sunlight was collected with the 70\,cm primary mirror of the VTT and imaged with a focal length of 45\,m to the LARS instrument. During the entire observation time, the seeing conditions were good and stable. Atmospheric distortions were actively corrected with the adaptive optics system \citep{2003SPIE.4853..187V} of the telescope. To generate the spatial context information for the spectroscopic measurement, a beamsplitter cube reflected 10\% of the incoming light to an context imager. A narrow-band interference filter was used to limit the light to the spectral G-band region at 4308\AA. The context camera recorded the images ($100\arcsec\times75\arcsec$) of the targeted sunspot region. To allow for post-facto image reconstruction, exposure times were set to around 15\,ms at a frame rate of 20\,Hz. Speckle reconstructions with the Kiepenheuer-Institute Speckle Interferometry Package \citep[KISIP,][]{2008SPIE.7019E..1EW} enabled observations at a spatial resolution approaching the diffraction limit of the telescope. For more information on the instrument specifications and the data calibration, we refer to the detailed descriptions by \citet{Doerr2015} and \citet{2017A&A...607A..12L}.

Figure \ref{fig_results_Ti5714_spot} shows the final context image of NOAA\,12146 observed on August 24 2014 at 8:41\,UTC. The sunspot umbra had a diameter of 15\arcsec\ and a magnetic field strength exceeding 2.5\,kG in the darkest core. Listed as Obs.\,8 in Table\,\ref{table_observations}, the sunspot center was located at a heliographic position of $\mathrm{X=361\arcsec}$ and $\mathrm{Y=36\arcsec}$, yielding a heliocentric angle of $\mathrm{\alpha=22.5^{\circ}}$. The context images of sunspots NOAA\,12150 (Obs.\,1), NOAA\,12149 (Obs.\,5), and NOAA\,12109 (Obs.\,13) are displayed in Figs.\,\ref{fig_results_Ti5714_spotA1}, \ref{fig_results_Ti5714_spotA2}, and \ref{fig_results_Ti5714_spotA3} in the appendix.

As indicated by the red circle in Figure \ref{fig_results_Ti5714_spot}, we selected the darkest part of the inner umbra for our spectroscopic analysis. Areas with identifiable umbral dots (UDs) were omitted. Since the light from the 3"-wide field of view is integrated for the spectroscopic measurements, and thus observed like a single point source, the spectral effect of any bright subarcsecond-sized UDs would not be distinguishable from the darker surroundings. At this point, we want to make a short excursion to the state of knowledge to explain why an inclusion of umbral dots would have falsified our velocity analysis. Previous studies have revealed a distinction between dots in the umbral center and dots at the umbral periphery close to the penumbra \citep{1986A&A...156..347G,2009ApJ...702.1048W}. While central UDs show either minor or slow upward flows of a few hundred $\mathrm{m\,s^{-1}}$, peripheral UDs feature upflows reaching up to $\mathrm{1000\,m\,s^{-1}}$ \citep[e.g.,][]{Schmidt+Balthasar1994,2004ApJ...614..448S,2004ApJ...604..906R,2007PASJ...59S.577K,2008ApJ...678L.157R,2009ApJ...694.1080S,2012ApJ...757...49W}. In line with the scenario of magneto-convection \citep{2006ApJ...641L..73S}, \citet{2010ApJ...713.1282O} and \citet{2013A&A...554A..53R} also found temporary returning downflows of several hundred $\mathrm{m\,s^{-1}}$ at the edges of some peripheral UDs. To comprehensively observe the presumably complex flow field of an umbral dot, a spatial resolution of around 0.1\arcsec\ would be required.

\subsection{Spectroscopic observations}
An extensive and comprehensive spectroscopic study was required to determine absolute velocities in sunspot umbrae with an accuracy of a few $\mathrm{m\,s^{-1}}$. Several important preconditions had to be fulfilled. To obtain a very good spectral precision, a spectrograph with a high resolution was essential. A laser frequency comb had to provide the absolute wavelength calibration. It also had to guarantee the absolute repeatability of the measurements and identical conditions for a direct spectral comparisons. To extract the average umbral velocities, a proper calibration of all systematic Doppler shifts from orbital, radial and rotational motions between the telescope and the observed sunspot umbra was mandatory. In addition to that, the reduction of the acoustic wave signal in umbrae required the observation of time series. Finally, the observed spectral line had to be ideally suited for Doppler shift measurements in the cool and dark environment of sunspot umbrae. To obtain Doppler velocities, the accurate reference wavelength of this line had to be provided.
In this section, we demonstrate how all of these requirements were met at an unprecedented scale.

\begin{figure}[htpb]
\includegraphics[trim=0cm 0cm 0cm 0cm,clip,width=\columnwidth]{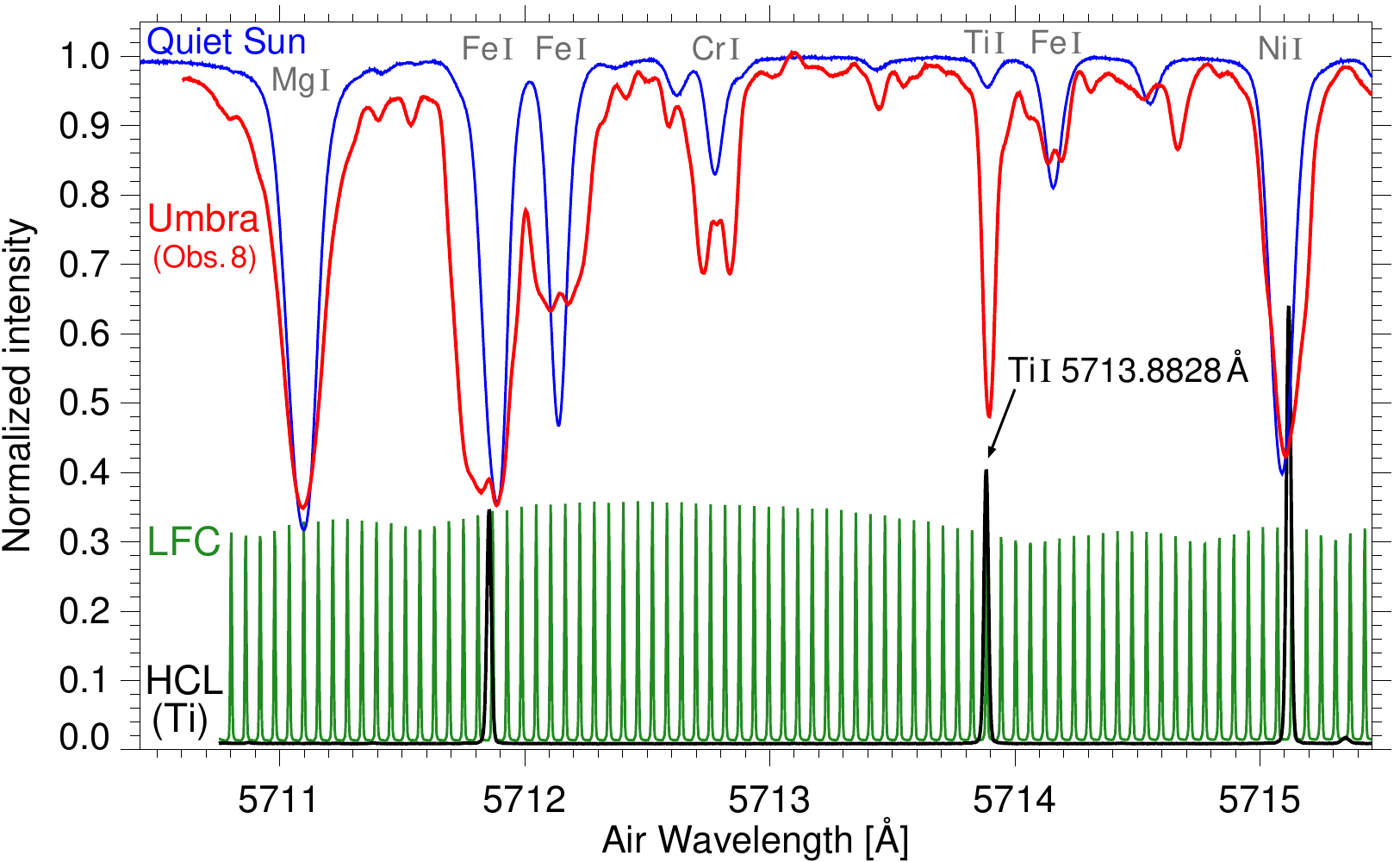}
\caption{Spectra observed with LARS in the wavelength region around 5713\,\AA. The solar absorption spectra were observed in the quiet Sun at disk center (blue curve), and in the sunspot umbra (red curve; based on Obs.\,8 in Table\,\ref{table_observations} and spatial context in Fig\,\ref{fig_results_Ti5714_spot}). The ions of the spectral lines are stated in gray color. The emission spectra of the laser frequency comb (LFC, green curve) and the titanium hollow cathode lamp (HCL, black curve) are required for the calibration of absolute wavelengths and Doppler shifts. The measured reference line of \ion{Ti}{I} at 5713.8828\,\AA\ is marked.}
\label{fig_results_Ti5714_profile}
\end{figure}

As discussed in the previous section, the beamsplitter of LARS reflects 10\% of the incoming sunlight to the context imager. The other 90\% are transmitted to the spectroscopic channel. A fiber-coupling unit {feeds} the light of a 3\arcsec\ field aperture (indicated by the red circle in Fig.\,\ref{fig_results_Ti5714_spot}) into a single mode fiber. The integrated solar signal {is} guided to a fiber switch. The other entrance ports of the fiber switch {are} linked to a laser frequency comb, a titanium hollow cathode lamp, and a tungsten lamp for flatfield calibration. The fiber switch changes the transmitted input channel. The output channel of the fiber switch {is} connected with a fiber to the entrance of VTT's echelle spectrograph. This guarantees an identical illumination of the echelle grating by all light sources and enables a valid calibration. At a wavelength of $\lambda=5714\,\AA$ and a full width at half maximum ($\mathrm{\Delta\lambda\approx8\,m\AA}$) of the instrumental profile, we obtain a spectral resolution of $\lambda/\Delta\lambda>700000$. The one-dimensional high-resolution spectrum {is} recorded with a charge-coupled device (CCD) camera. The 5\,\AA\ wide spectral range around 5713\,\AA\ was centered along the 2048\,pixel chip size of the camera. Figure \ref{fig_results_Ti5714_profile} shows an overview of the solar absorption spectra recorded in a quiet Sun region at disk center and in a sunspot umbra (here NOAA\,12146), as well as the artificial emission spectra of the laser frequency comb and the titanium hollow cathode lamp.

{A} laser frequency comb served for the calibration of the spectrograph and the determination of the absolute wavelength scale of the solar spectrum. A pulsed femtosecond laser generated an emission spectrum with a mode separation of exactly 5.445\,GHz (or $\mathrm{59.3\,m\AA}$\ at $\mathrm{\lambda=5713\,\AA}$). A more detailed descriptions of the instrument specifications was given by \citet{Doerr2015}. The light level of the frequency comb was adjusted to the intensity of the solar spectrum. The final comb spectrum is displayed in Fig.\,\ref{fig_results_Ti5714_profile} at one third of its original intensity level.

The solar spectrum and the frequency comb spectrum were observed in an alternating order, enabling a quasi-simultaneous wavelength calibration. Thus, each observation cycle consisted of one solar spectrum and one frequency comb spectrum. The exposure time of the camera was set to 1\,s (for Obs.\,1--11, and to 0.9\,s for Obs.\,12--13 in Table\,\ref{table_observations}). Given the required exposure time for both spectra, plus a systematic offset time, we obtained a total cycle time of 2.5\,s (for Obs.\,1--11, and to 2.0\,s for Obs.\,12--13 in Table\,\ref{table_observations}). With a temporal cadence at the second-scale, we were able to effectively correct for drifts of the spectrograph. The analysis of convective shifts in sunspot umbrae demanded the observation of time series which were long enough to reduce the signal of acoustic waves. Since oscillations with periods around 5\,min dominate the umbral activity, we observed time sequences of typically around 30\,min. The total number of cycles and the observation time of each sequence are listed in Table\,\ref{table_observations}. Each sunspot was observed at least twice, thus for at least 60\,min. In total, we observed 13 sequences in four different sunspots. The sum of all observations made a time of 475\,min, or almost 8\,h. 

Using the data pipeline developed by \citet{Doerr2015}, a careful data calibration was performed in order to exploit the unprecedented spectral accuracy of our observations. Using the white light spectrum from a tungsten lamp, we applied a spectral flatfield correction to all data. This included the elimination of background noise, intensity gradients and camera defects. The absolute wavelength calibration was done with the emission spectrum of the laser frequency comb. The relative wavelength calibration is given by the constant mode spacing and fixed offset frequency. To obtain the absolute scale, it was sufficient to identify the frequency (and thus the wavelength) of a single mode by the proximity of a well-known solar spectral line as the reference for an unambiguous wavelength calibration. Each single solar spectrum of an observation sequence was calibrated by the interpolation between the preceding and subsequent comb spectrum. The absolute wavelength calibration is accurate at the level of a few femtometer, which translates into a velocity accuracy of $\mathrm{\sigma_{comb}=1\,m\,s^{-1}}$ \citep{Doerr2015}. In the next step, we reduced all inherent systematic Doppler shifts originating from time-dependent relative motions between the telescope and the observed sunspot umbra. These included all orbital, radial and rotational motions of the Sun and Earth, with a given uncertainty $\mathrm{\sigma_{eph}}$ of around $\mathrm{0.1\,m\,s^{-1}}$. We used the ephemerides code developed by \citet{Doerr2015}, which in turn was based on NASA's SPICE toolkit \citep{Acton1996}. The rotation of the Sun at the observed heliographic position was computed according to the spectroscopic differential rotation model by \citet{1990ApJ...351..309S}. With a telescope pointing accuracy of 1\arcsec\ and the estimated uncertainty of the model, we set the velocity error caused by solar rotation to $\mathrm{\sigma_{rot}=4\,m\,s^{-1}}$. Finally, the solar spectra contained only line shifts caused by local motions in the umbra itself, and the constant gravitational redshift of $\mathrm{+633\,m\,s^{-1}}$ caused by the Sun and Earth according to the principle of equivalence and the theory of general relativity. More information on the data calibration was given by \citet{Doerr2015}, a schematic overview can be found in \citet{2017A&A...607A..12L}.

\begin{figure}[htpb]
\includegraphics[trim=0cm 0cm 0cm 0cm,clip,width=0.88\columnwidth]{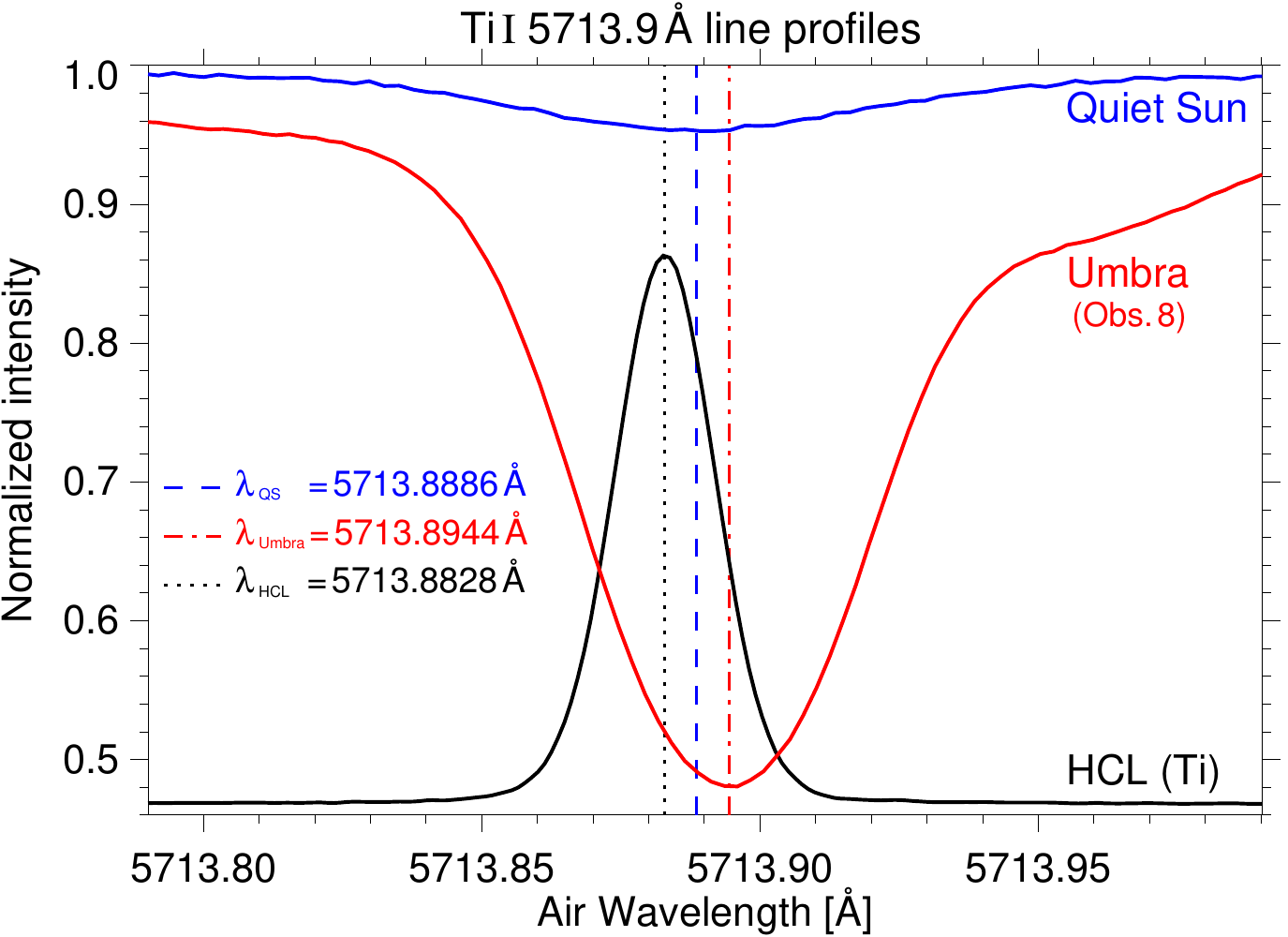}
\caption{High-resolution observations of the \ion{Ti}{i}\,5713.9\,\AA\ line as a detailed view of Fig.\,\ref{fig_results_Ti5714_profile}. The quiet Sun spectrum (blue curve) and umbral spectrum (red curve) are compared with the emission spectrum from the hollow cathode lamp (HCL, black curve). While both solar spectra were normalized to the continuum level, the lamp spectrum was adapted in intensity to allow a better illustration. The center positions of fits to the line core are marked as blue dashed (quiet Sun), red dashed-dotted (umbra), and black dotted (lamp) lines.}
\label{fig_results_Ti5714_line}
\end{figure}

We focused our study of umbral velocities on the \ion{Ti}{I} line at 5713.9\,\AA. When comparing the spectral lines of the observed wavelength range (Fig.\,\ref{fig_results_Ti5714_profile}), it becomes apparent that in the cool magnetized atmosphere of a sunspot umbra most lines split or are blended by other atomic and molecular lines. The \ion{Ti}{I} line constitutes a rare exception. With a Land\'e factor of $\mathrm{g=0}$, it is magnetically insensitive and therefore does not show a Zeeman splitting in the strongly magnetized umbra. The line becomes stronger in the cool umbral atmosphere. With respect to the spectral continuum, it increases in line depth from around 5\% in the quiet Sun to a line depth of around 50\% in the umbra. The line core and inner wings (of at least $\mathrm{\pm25\,m\AA}$ from the line center) are not affected by line blends. Thus, the \ion{Ti}{I} 5713.9\,\AA\ line is best suited for umbral measurements focussing on Doppler shifts and bisector profiles. Conveniently, the results of the spectroscopic analysis allow for a direct comparison with the results of \citet{Beckers1977}. A detailed view of the spectral line is displayed in Fig.\,\ref{fig_results_Ti5714_line}.

A crucial step to increase the accuracy of the obtained Doppler velocities was the measurement of the laboratory wavelength of \ion{Ti}{I} 5713.9\,\AA\ with LARS. We recorded the spectrum of a titanium hollow cathode lamp and calibrated the absolute wavelengths with the frequency comb. The entire spectrum is shown in Fig.\,\ref{fig_results_Ti5714_profile}. The detailed view of the induced emission line at 5713.9\,\AA\ is plotted in Fig.\,\ref{fig_results_Ti5714_line} at an adapted intensity scale. The very good approximation of the line profile with a Voigt function yielded a central wavelength of $\mathrm{\lambda_0=5713.8828\AA}$ with an error of $\mathrm{0.05\,m\AA}$. At the given wavelength, the error for the wavelength reference translates into an uncertainty of $\mathrm{\sigma_{\lambda_0}=2.6\,m\,s^{-1}}$ for the calculation of Doppler velocities. In comparison, the observed reference wavelength provided by the National Institute of Standards and Technology Atomic Spectra Database \citep[NIST ASD,][]{NIST_ASD} is stated as 5713.881\,\AA\ with an uncertainty of $\mathrm{4\,m\AA}$ \citep{1991PhyS...44..446F}. Thus, we have increased the accuracy of the reference wavelength by almost two orders of magnitude. At this point, we note that throughout the paper we adopted common practice in astrophysics and used air wavelengths when referring to the wavelength of spectral lines.

Knowing the laboratory wavelength $\mathrm{\lambda_0}$ (or $\mathrm{\lambda_{HCL}}$) of the spectral line, it was possible to calculate the Doppler shift of the observed wavelength position $\lambda$ of the solar spectral line. The Doppler velocity $\mathrm{v_{los}}$ was determined by
\begin{equation}
\mathrm{v_{los}=c\cdot(\lambda-\lambda_0)/\lambda_0-v_{grs}}\label{eq1}
\end{equation}
in which $\mathrm{c}$ was the speed of light. The gravitational redshift of $\mathrm{v_{grs}=+633\,m\,s^{-1}}$ (or  12.1\,m\AA\ at $\mathrm{\lambda=5713.9\,\AA}$) was subtracted. As clearly recognizable in Fig.\,\ref{fig_results_Ti5714_line}, the gravitational redshift is still included in the solar line profiles. The systematic observational uncertainty $\mathrm{\sigma_{syst}=(\sigma^2_{comb}+\sigma^2_{eph}+\sigma^2_{rot}+\sigma^2_{\lambda_0}})^{1/2}$ was given by the accuracies of the laser frequency comb, the ephemeris correction, the telescope pointing and solar rotational, and the spectral line's wavelength reference. We thus obtained a total systematic uncertainty of $\mathrm{4.9\,m\,s^{-1}}$  for the resultant Doppler velocities.

\section{Results and discussion}\label{sec3_results}
In this section, we will highlight the fundamental results and novel findings obtained with our absolute velocity measurements in sunspot umbrae. The observations presented in this work excel by an unprecedented accuracy. Therefore, our main goal was to determine the absolute vertical velocity in the cores of sunspot umbrae. The shape of the umbral \ion{Ti}{I} 5713.9\,\AA\ line profile in Fig.\,\ref{fig_results_Ti5714_line} motivates a spectral analysis by means of two techniques. Firstly, fitting a 50\,m\AA\ wide symmetrical Gaussian function to the spectral line core provided a resilient line center wavelength (here $\mathrm{\lambda_{Umbra}=5713.8944\,\AA}$) and the average Doppler shift. Secondly, performing a bisector analysis of the line profile revealed the noticeable asymmetry of the outer line wings. Further methodological details and the main results are discussed in Section \ref{sec3_velocities} and \ref{sec3_asymmetry}. By briefly illustrating the temporal variation of umbral velocities during a sequence in Section \ref{sec3_temporal}, we do not want to distract from the key results. We rather want to clarify that temporal averaging is mandatory in order to discern the convective blueshift of an umbra. In Section \ref{sec3_dependences}, we perform a statistical analysis on several important parameters, compute their correlations, and give the observational verification of significant dependences in sunspot umbrae.

\subsection{Temporal variation}\label{sec3_temporal}
Similar to acoustic wave modes in the quiet Sun, sunspot umbrae exhibit a wealth of magneto-acoustic waves. In the umbral photosphere, the mixture of modes is dominated by waves with periods around 5\,min \citep{1984ApJ...285..368T}. In our time sequences, these oscillations in Doppler shift superimpose the global convective shift in umbrae. An exemplary time series of line center oscillations is shown in Fig.\,\ref{fig_results_Ti5714_oscillation} in the appendix. The amplitudes reach Doppler shifts of up to 5\,m\AA, or around $\mathrm{260\,m\,s^{-1}}$. The standard deviation for a single measurement of this explicit sequence amounts to $\mathrm{87\,m\,s^{-1}}$. For all other time series, the standard deviation ranges from $\mathrm{40\,m\,s^{-1}}$ to $\mathrm{100\,m\,s^{-1}}$. Proper temporal averaging reduces the wave signal and yields the mean wavelength position. The uncertainty of the mean decreases substantially to a few $\mathrm{m\,s^{-1}}$, given by the division of the standard deviation by the square root of the number of observation cycles. By applying Eq.\,\ref{eq1}, the mean convective Doppler shift of our exemplary case yields a blueshift of $\mathrm{-21.5\,m\,s^{-1}}$. For each observation sequence, the temporally averaged velocities are listed in the two right columns of Table\,\ref{table_observations}. The respective uncertainty of the mean velocity is listed in brackets and includes all systematic and statistic errors according to the propagation of errors.

As a byproduct, we performed a Fourier analysis of the wave power of the combined observation sequences. The power spectrum (see Fig.\,\ref{fig_results_frequency_analysis} in the appendix) displays several distinct peaks at periods between 150\,s and 380\,s. Dominating wave periods at different scales can be identified at 150\,s, 195\,s, 240\,s, 280\,s, and 330\,s. The mixture of magneto-acoustic wave modes is distinctive for the umbral photosphere below the acoustic cut-off layer. At the line formation height of \ion{Ti}{i}\,5713.9\,\AA, all acoustic wave modes are present. At higher atmospheric layers, the acoustic cut-off period of around 192\,s would limit the upward propagation along the magnetic field lines to waves with shorter periods \citep{2016PhDT........15L}.

We note that from here on, we proceed our study with the interpretation of the temporally averaged line profiles. 

\subsection{Absolute velocities}\label{sec3_velocities}
The averaged line profiles of an umbra and the quiet Sun at disk center are displayed in Fig.\,\ref{fig_results_Ti5714_line}. To obtain the central wavelength position of the spectral line, we applied a symmetrical Gaussian fit to the line core. The observed line minimum served as the center of the 50\,m\AA\ ($\mathrm{\pm25\,m\AA}$) wide range of the fit function. Thus, the fit accounts only for the lower third of the line profile which makes the largely symmetrical line core. We identify the center position of the Gaussian function as the line center wavelength. 
For our example, we obtain a wavelength $\mathrm{\lambda_{QS}=5713.8886\,\AA}$ for the quiet Sun and $\mathrm{\lambda_{Umbra}=5713.8944\,\AA}$ for the umbra. The shift of the line center is demonstrated by vertical lines in Fig.\,\ref{fig_results_Ti5714_line}. 
The translation of air wavelength into Doppler shifts (Eq.\,\ref{eq1}) yields Doppler velocities of $\mathrm{-328.7\,m\,s^{-1}}$ for the quiet Sun and $\mathrm{-21.5\,m\,s^{-1}}$ for the umbra. Consequently, the convective blueshift has decreased substantially to a residual of $6\%$ of the blueshift in the quiet Sun. 

The thus obtained umbral velocities are listed as $\mathrm{v_{los, \pm25\,m\AA}}$ in the right column of Table\,\ref{table_observations} for each observation sequence. For observations 1--11, the convective blueshifts range between $\mathrm{-61.8\,m\,s^{-1}}$ and $\mathrm{-15.7\,m\,s^{-1}}$. The average Doppler velocity for these three sunspots (NOAA\,12150, 12149, and 12146) amounts to $\mathrm{-29.8\,m\,s^{-1}}$. In comparison with the quiet Sun, the convective blueshift has thus decreased by around $\mathrm{300\,m\,s^{-1}}$, or less than 10\% compared to the quiet Sun. Separated into sunspots, NOAA\,12150 shows the strongest blueshift of around $\mathrm{-51.5\,m\,s^{-1}}$ (mean of Obs.\,1--3). NOAA\,12146 yields the weakest blueshift of $\mathrm{-23.4\,m\,s^{-1}}$ (mean of Obs.\,6--11). 

The measurements of NOAA\,12109 (Obs.\,12--13) constitute an apparent exception from the general trend (also discussed at the end of Section\,\ref{sec3_asymmetry}). In contrary to all other sunspots, the umbra features positive Doppler velocities of around $\mathrm{+44.2\,m\,s^{-1}}$ on average. Including these two measurements in the calculation of the overall umbral Doppler velocity, we still receive a blueshift of $\mathrm{-20.5\,m\,s^{-1}}$. Our accurate velocity measurements verify the finding of either negligible or moderate blueshifts in umbrae \citep[e.g.,][]{1984SoPh...93...53K,Schmidt+Balthasar1994,2004A&A...427..319B,2004A&A...415..717T,2006A&A...454..975R}.

In order to relate our results with the work of \citet{Beckers1977}, a subtle adaption of the spectral analysis is crucial. So far, we have obtained the spectral line center by a Gaussian fit with a width of 50\,m\AA\ ($\mathrm{\pm25\,m\AA}$ around the line minimum). In contrast, \citet{Beckers1977} applied a 100\,m\AA\ ($\mathrm{\pm50\,m\AA}$) wide Gaussian function to the \ion{Ti}{i}\,5713.9\,\AA\ line. To allow for a direct comparison, we therefore repeated our spectral analysis from above, but with the fit range of 100\,m\AA. Regarding the umbral profile in Fig.\,\ref{fig_results_Ti5714_line}, the fit accounts for the lower two thirds of the line profile. This implies that the fit range partially includes the bend of the red line wing. As a consequence, the spectral analysis is affected by the asymmetry of the \ion{Ti}{i}\,5713.9\,\AA\ line.
The resultant Doppler velocities are listed as $\mathrm{v_{los, \pm50\,m\AA}}$ in the second right column of Table\,\ref{table_observations}. With respect to the velocities $\mathrm{v_{los, \pm25\,m\AA}}$, the velocities $\mathrm{v_{los, \pm50\,m\AA}}$ are shifted by around $\mathrm{+25\,m\,s^{-1}}$ to the red. The average Doppler velocity for observations 1 to 11 (see Table\,\ref{table_observations}) then amounts to $\mathrm{-5.2\,m\,s^{-1}}$. The average for all observations is $\mathrm{+6.3\,m\,s^{-1}}$. These velocities around zero are in very good agreement with the results of \citet{Beckers1977}, on which he based his conclusion of umbrae being at rest in the absence of convection. Based on our new findings, this statement has to be qualified.

\subsection{Bisector analysis}\label{sec3_asymmetry}
In the previous section, we have seen that the asymmetric outer line wings disturb the spectral analysis. Thus, we performed a bisector analysis for a closer inspection of the spectral line asymmetry and the consequential height distribution of Doppler velocities in the umbral atmosphere. A bisector describes the connection of the midpoints of a number of horizontal line segments at distinct line depths. We performed the analysis on the temporally averaged line profile of each observation sequence. For umbral line profiles like in Fig.\,\ref{fig_results_Ti5714_line}, bisector positions were calculated from the line minimum to an upper threshold of 90\% of the continuum intensity.  For weaker quiet Sun profiles, the upper level was set to 99\%. For umbral bisectors, we sampled up to 15 equidistant positions with a step size of around 3\% of the continuum intensity. In the following, we exemplarily discuss the bisector curves of the \ion{Ti}{i}\,5713.9\,\AA\ line profiles shown in Fig.\,\ref{fig_results_Ti5714_line}.

\begin{figure}[htpb]
\includegraphics[trim=0cm 0cm 0.0cm 0cm,clip,width=\columnwidth]{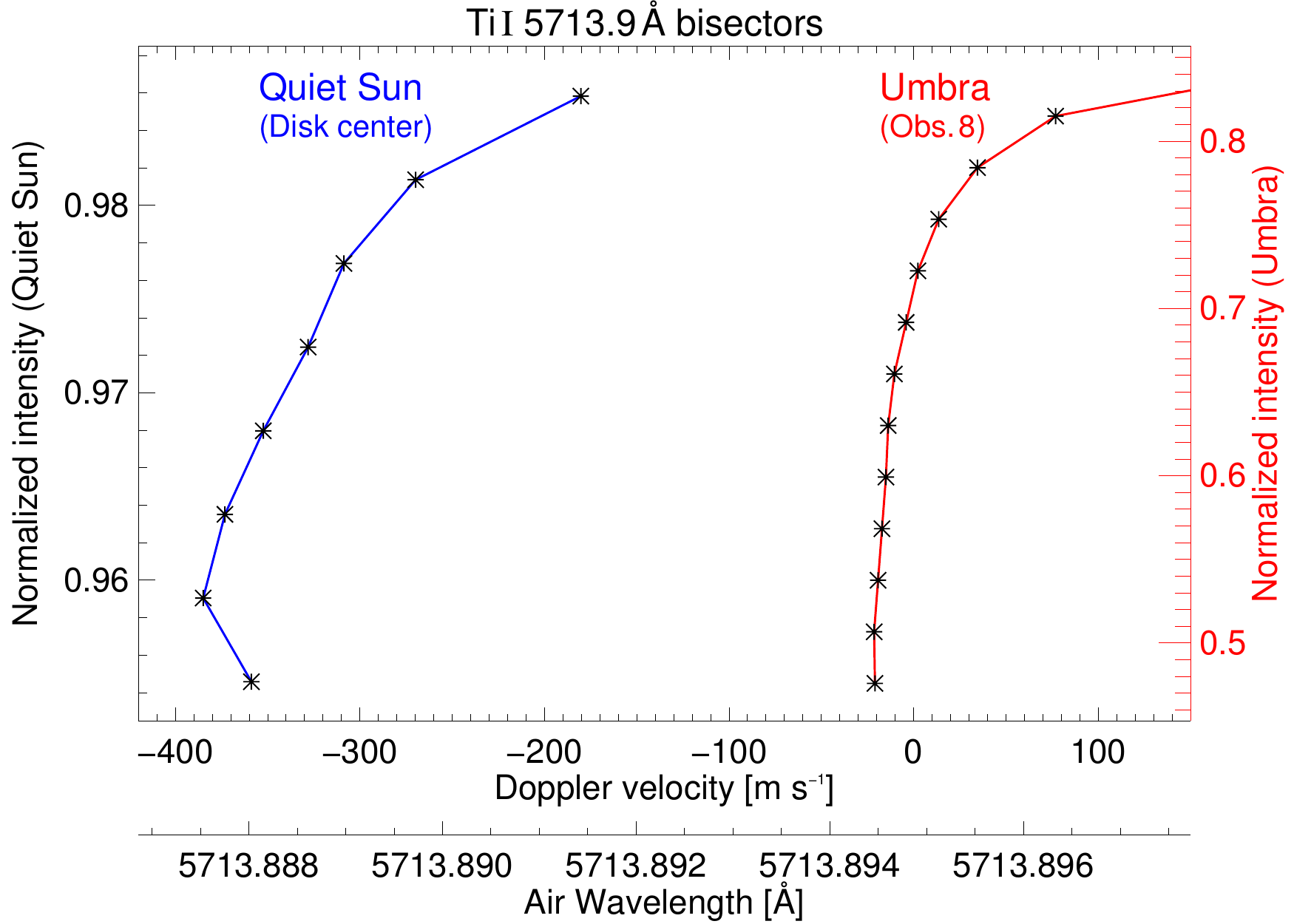}
\caption{Bisector analysis of the \ion{Ti}{i}\,5713.9\,\AA\ line profiles shown in Fig.\,\ref{fig_results_Ti5714_line}. The bisector curves for the quiet Sun (blue curve) and for the sunspot umbra (red curve) are displayed at the absolute wavelength scale and in Doppler velocity. The normalized intensities are given at two different y-axis scales to allow a better visual comparison.}
\label{fig_results_Ti5714_bisector}
\end{figure}

Figure \ref{fig_results_Ti5714_bisector} displays the bisector curves of the quiet Sun and the umbra. The curves are given in air wavelength and in Doppler velocity according to Eq.\,\ref{eq1}. Two major differences are apparent for the position and shape of both curves. Regarding the position, the quiet Sun bisector reaches convective blueshifts of up to $\mathrm{-380\,m\,s^{-1}}$. The umbral bisector yields a velocity of around $\mathrm{-20\,m\,s^{-1}}$ only. Concerning the shape, the quiet Sun bisector indicates the differential convex asymmetry resembling a C-shape \citep{1981A&A....96..345D} which is typical for photospheric lines in the presence of unresolved convective motions. Toward the line minimum, the bisector profiles start to reverse in their convective blueshift. On the other hand, the bisector curve of the umbra does not show such a trend. Admittedly, it demonstrates a similar shift toward longer wavelength when the curve approaches the continuum intensity, but this is due to the line blend in the outer right wing of the \ion{Ti}{i}\,5713.9\,\AA\ line. In the lower half of the line, the bisector profiles rather  saturates and becomes a straight line which suggests a negligible or very low amount of convection. With regard to our velocity analysis in Section \ref{sec3_velocities}, we argue that the Gaussian fit to the lower half of the spectral line yields robust Doppler velocities $\mathrm{v_{los, \pm25\,m\AA}}$.

\begin{figure}[htpb]
\includegraphics[trim=0.1cm 0cm 0.4cm 0cm,clip,width=\columnwidth]{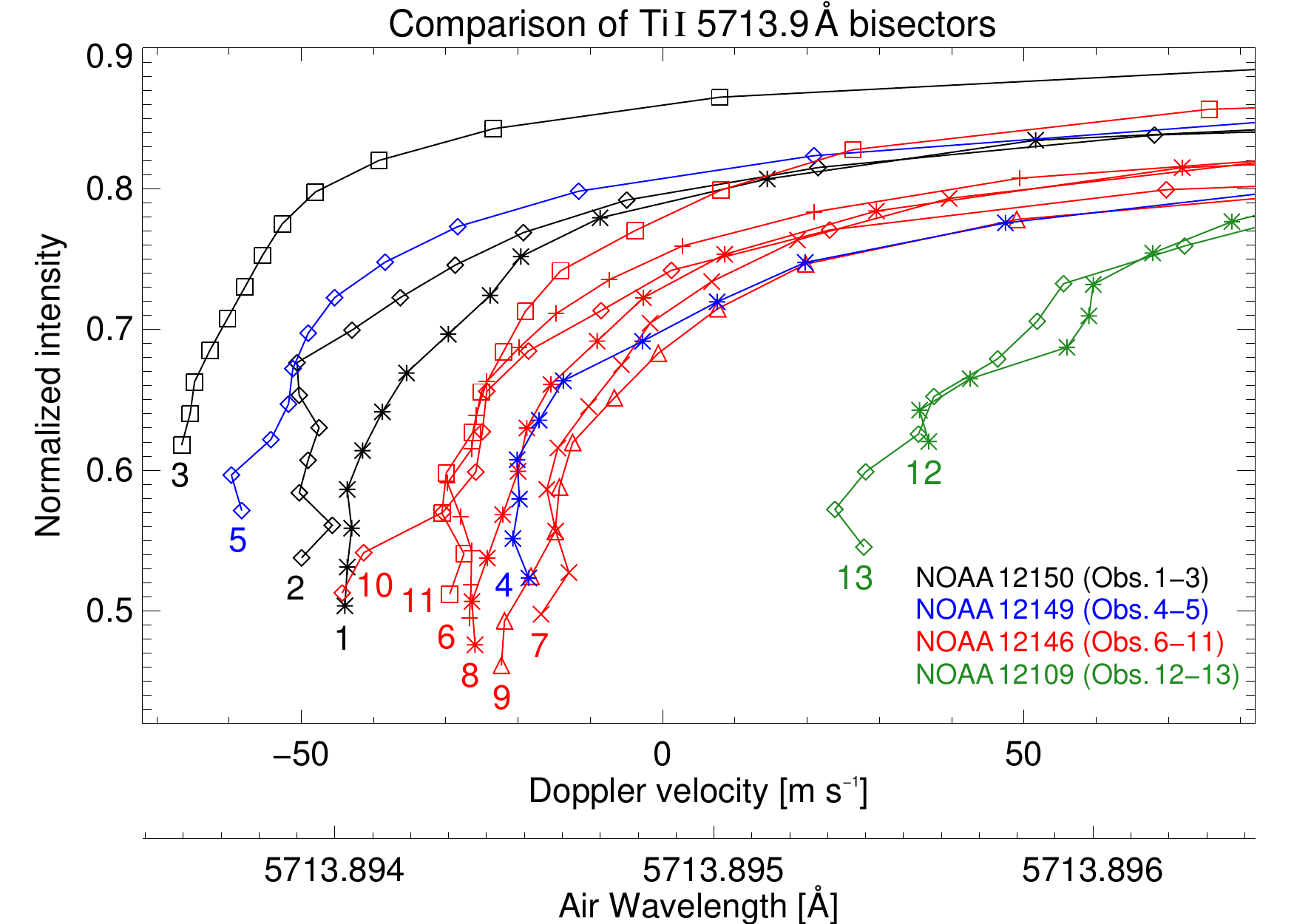}
\caption{Bisector curves of \ion{Ti}{i}\,5713.9\,\AA\ for all 13 observation sequences (see Table\,\ref{table_observations}) in comparison. Bisectors observed for the same sunspot are displayed in the same color. The curves are given in absolute wavelengths and Doppler velocities.}
\label{fig_results_Ti5714_bisector_comparison}
\end{figure}

Subsequently, we have performed the described bisector analysis for all umbral observations. The average bisector curves of all 13 sequences (as listed in Table\,\ref{table_observations}) are compared in Fig.\,\ref{fig_results_Ti5714_bisector_comparison}. Apart from offsets in Doppler velocity and in intensity, all bisector curves of observations 1--11 feature the same trend. Toward the line minimum, the curves become almost straight vertical lines. Thus, the line core is symmetrical. Line minima arrive at blueshifts between $\mathrm{-67\,m\,s^{-1}}$ (Obs.\,3) and $\mathrm{-17\,m\,s^{-1}}$ (Obs.\,7), and normalized intensities between 0.46 (Obs.\,9) and 0.62 (Obs.\,3). Observations 12 and 13 form an exception. The bisector curves present a stronger offset in Doppler velocity and a steeper gradient lacking the usual saturation to a vertical straight line. 

We can only speculate about the reasons for the discrepancy of observations 12 and 13. Observationally, the atmospheric seeing conditions were not as good as for the other observations. This could have produced a possibly higher amount of stray light affecting the spectral profile which would explain the weaker line depth. But it could not explain the redshift. On the contrary, a larger amount of light from the surrounding quiet Sun would cause an additional convective blueshift to the line profile. An other scenario implies the location of the sunspot. At a heliocentric angle of around 60\%, the umbra is closer to the solar limb than the other umbrae. At this scale, line-of-sight effects can become important. The observation of a sunspot umbra with a large line-of-sight angle could involve the partial measurement of penumbral flow fields. Moreover, with increasing heliocentric angle spectral lines sample higher atmospheric layers which could feature other complex atmospheric motions. But contradicting to this scenario, our observed umbral velocities have not shown any significant dependence on the heliocentric angle. The likeliest explanations for the umbral redshift arise from the sunspot itself. The umbra of NOAA\,12109 is larger than the others and has a stronger magnetic field strength (see Table\,\ref{table_observations}). A complete absence of convection would indeed lead to a decrease of the blueshift. To verify these connections, we aim to perform additional observations in the future and to carry out a more quantitative study. For the following analysis of the dependences of different umbral parameters, we will leave observations 12 and 13 out of the discussion.

\subsection{Dependence of umbral velocities on other variables}\label{sec3_dependences}
The distribution of bisectors in Fig.\,\ref{fig_results_Ti5714_bisector_comparison} (Obs.\,1--11) suggests an apparent dependence of the Doppler shifts on the intensity of the spectral line minima, i.e., the line strength. The blueshift decreases with increasing line depth. To verify this and potentially further relations, we applied a statistical analysis to our data. A normality test \citep{normality_test} on the data of observations 1--11 confirmed the normal distribution of the studied parameters. These are the heliocentric angle $\alpha$, the line minimum intensity $\mathrm{I_{min}}$, the line-of-sight velocity $\mathrm{v_{los}=v_{los, \pm25\,m\AA}}$, the projected vertical velocity component $\mathrm{v_{z}=v_{los}/\cos{\alpha}}$, and the magnetic field strength $\mathrm{B_{abs}}$ (as in Table \ref{table_observations}). On this basis, we computed the linear correlation coefficients \citep{Pearson253} between the normally distributed variables. A two-sided $t$-test returned the probability value $p$ of the correlation. We interpret $p$-values smaller than $0.05$ to be an indication for a high significance of the correlation. Table \ref{table_correlations} lists the coefficients $\rho$ and $p$-values for the correlations between the variables x and y. 

\begin{table}[htbp]
\caption{Correlation coefficients $\rho$ between variables x and y.}
\label{table_correlations}
\centering
\begin{tabular}{cccc}
\hline\hline
x&y&$\rho$(x,y)&$p$\\ 
\hline 
$\alpha$&$\mathrm{I_{min}}$&$+0.34$&0.3073\\
$\alpha$&$\mathrm{v_{los}}$&$-0.39$&0.2399\\
$\alpha$&$\mathrm{v_{z}}$&$-0.58$&0.0598\\
$\mathrm{I_{min}}$&$\mathrm{v_{los}}$&$-0.85$&0.0008\\
$\mathrm{I_{min}}$&$\mathrm{v_{z}}$&$-0.86$&0.0007\\
$\mathrm{B_{abs}}$&$\mathrm{v_{z}}$&$+0.79$&0.0039\\
\hline
\end{tabular}
\tablefoot{Small probability values ($p<0.05$) indicate high significance.}
\end{table}

With correlations coefficients between $+0.34$ and $-0.58$, and $p$-values larger than $0.05$, we find no significant linear dependence on the heliocentric angle $\alpha$. Neither the line minimum intensity $\rm I_{min}$ nor the observed umbral velocities $\rm v_{los}$ and $\rm v_{z}$ seem to vary with the position of the sunspot on the solar disk. On the contrary, the Doppler velocity shows a significant dependence on the line minimum intensity. The correlation coefficient is $\mathrm{\rho(I_{min},v_{los})}=-0.85$ with a very small $p$-value. Evidently, the convective blueshift of the umbra decreases for increasing line depth, or decreasing line minimum intensity. For the projected vertical velocity component $\mathrm{v_{z}}$, we obtain the same, even more significant, dependence on the spectral line depth. The correlation coefficient is $\mathrm{\rho(I_{min},v_{z})}=-0.86$ with a very small $p$-value. This relation makes sense since the line depth represents a measure for the umbral temperature. To be exact, the depth of the \ion{Ti}{i}\,5713.9\,\AA\ line increases in the cool environment of the strongly magnetized umbral atmosphere. The decline of convection leads to a reduced heat flux, a cooler atmosphere, a deeper \ion{Ti}{i} line, and of course a reduction of the convective blueshift. In this context, it is reasonable that the umbral velocity displays a significant positive correlation of $\mathrm{\rho(B_{abs},v_{z})}=+0.79$ with the magnetic field strength. It is generally accepted \citep{1941VAG....76..194B,1965ApJ...141..548D}, that a stronger magnetic field leads to a more effective suppression of convective heat flux, and thus reduced convective blueshift. To our best knowledge, this is the first time that this full chain of associations has been proven observationally for sunspot umbrae.

\begin{figure}[htpb]
\includegraphics[trim=0.0cm 0cm 0.0cm 0cm,clip,width=\columnwidth]{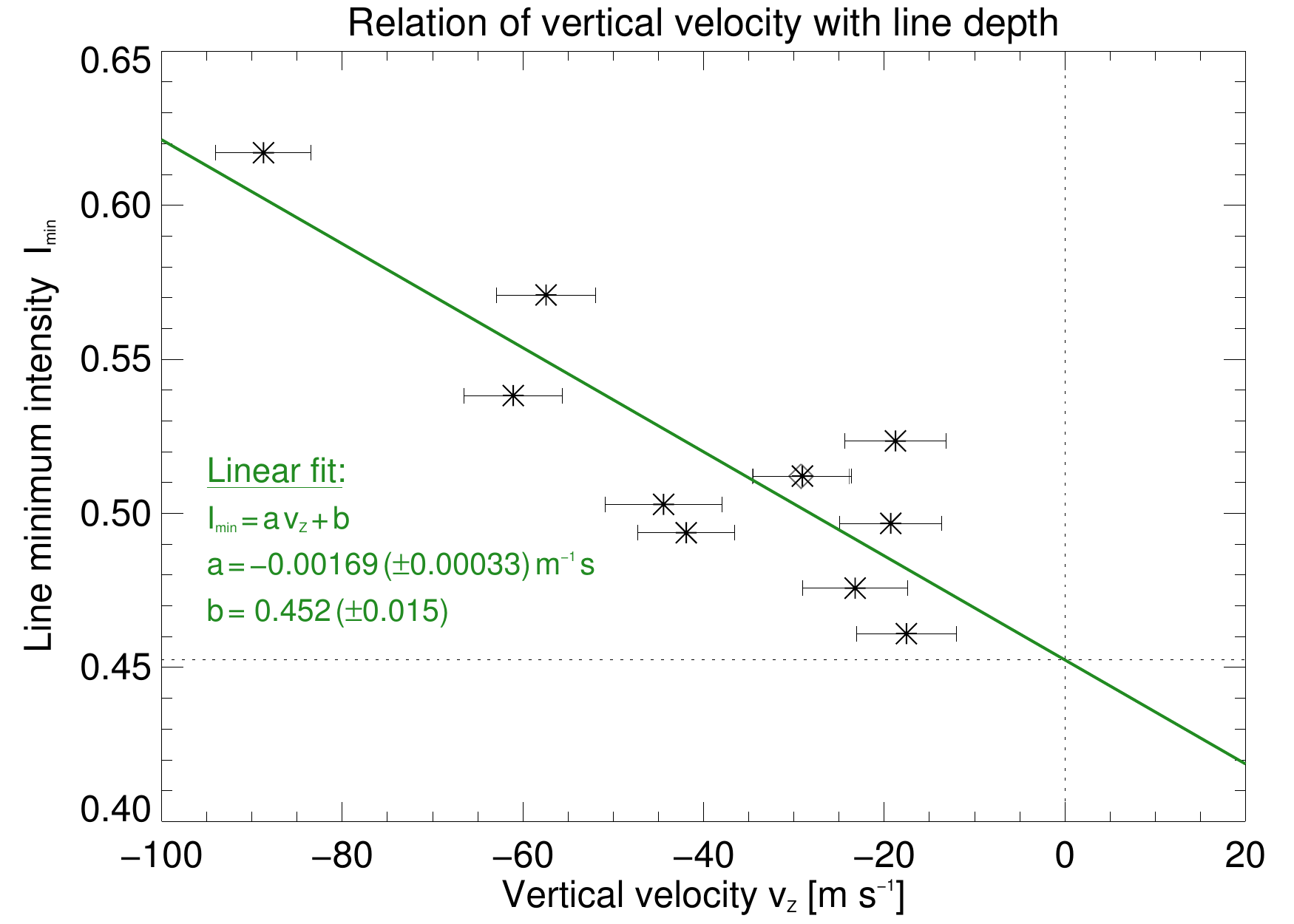}
\caption{Normalized line minimum intensities $\mathrm{I_{min}}$ plotted against the vertical velocities $\mathrm{v_{z}}$. The measurements (black asterisks) are approximated by a linear function (green line) with slope a and intercept b, and their respective errors in brackets.}
\label{fig_results_fit_vz_Imin}
\end{figure}

\begin{figure}[htpb]
\includegraphics[trim=0.0cm 0cm 0.0cm 0cm,clip,width=\columnwidth]{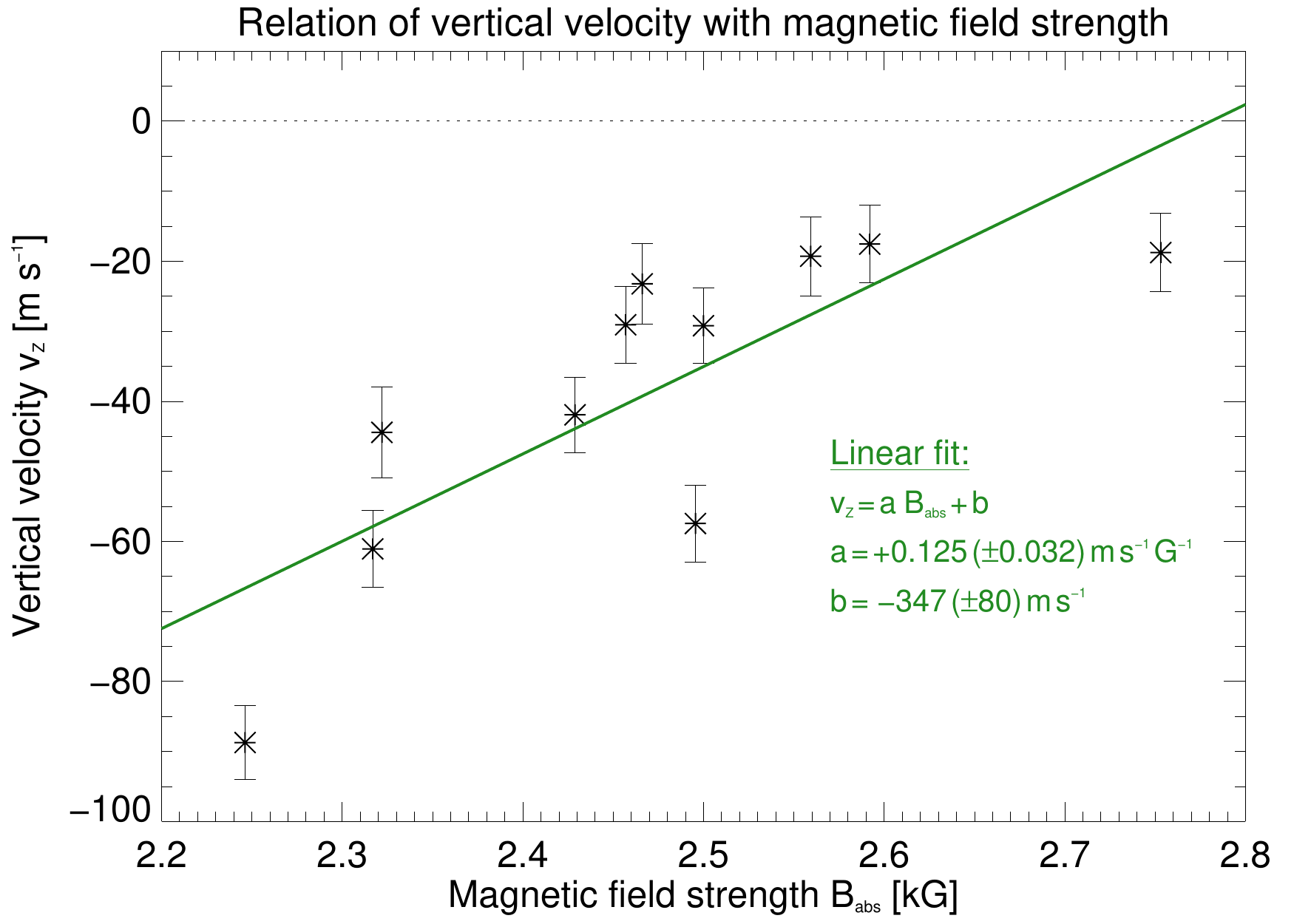}
\caption{Vertical velocities $\mathrm{v_{z}}$ plotted against the absolute magnetic field strengths $\mathrm{B_{abs}}$ (in $10^3$ Gauss). The measurements (black asterisks) are approximated by a linear function (green line) with slope a and intercept b, and their respective errors in brackets.}
\label{fig_results_fit_Babs_vz}
\end{figure}

We illustrate the dependence of the umbral velocity on the line minimum intensity and on the magnetic field strength in Figs.\,\ref{fig_results_fit_vz_Imin} and \ref{fig_results_fit_Babs_vz}, respectively. In both cases, the distribution of points can be interpreted by a linear trend which is specified by the linear fit functions. The values of the slopes a and y-intercepts b are stated in the plots with their respective errors. In Fig.\,\ref{fig_results_fit_vz_Imin}, the function decreases from line minimum intensities of 0.62 at velocities of $\mathrm{-100\,m\,s^{-1}}$ to intensities of about 0.45 at zero velocity. In Fig.\,\ref{fig_results_fit_Babs_vz}, the curve runs from velocities of $\mathrm{-72\,m\,s^{-1}}$ at magnetic field strength of 2.2\,kG to the zero velocity point at field strength of 2.78\,kG. According to that, a regular sunspot umbra with a field strength of 2.5\,kG would result in a convective blueshift of $\mathrm{-35\,m\,s^{-1}}$ at a line minimum intensity of 0.51. An entire absence of convection and zero velocity level would be reached for sunspots with magnetic field strength of 2.78\,kG. Vice versa, sunspot or pores with weaker magnetic fields feature stronger blueshifts as suggested by \citet{2006A&A...454..975R}. Despite the fact that our sample of observations is relatively small, we interpret our results and confirmed relationships as significant and robust. We note that the extrapolation of the fit in Fig.\,\ref{fig_results_fit_Babs_vz} to zero magnetic field strength yields a velocity of $\mathrm{-347\,m\,s^{-1}}$. This corresponds very well with the observed convective blueshift of the quiet Sun (compare Fig.\,\ref{fig_results_Ti5714_bisector}). This conformity substantiates the continuity of linear relation between the magnetic field strength and the convective blueshift. It remains to be tested whether solar magnetic features with intermediate field strength reveal consistent blueshifts. To test the consistency of the relation between the line depth and the magnetic field strength, we insert the line minimum intensity of 0.96 into the extrapolated linear function in Fig.\,\ref{fig_results_fit_vz_Imin}. We yield an umbral velocity of $\mathrm{-300\,m\,s^{-1}}$ which is of the order of the observed convective blueshift of the quiet Sun. Inserted into the function in Fig.\,\ref{fig_results_fit_Babs_vz}, the line depth of the quiet Sun profile is attained by a magnetic field strength of 370\,G. Within the error margins of the fits, the modeled linear relations between the magnetic field strength, the convective blueshift, and the spectral line depth are in good agreement.

which is compatible within the uncvertainties (was sind die denn) to the observed

\section{Summary and conclusions}\label{sec4_conclusions}
We find that the common assumption by \citet{Beckers1977} of sunspot umbrae being at rest by the absence of convection has to be revised. We measured convective blueshift of a few tens of $\mathrm{m\,s^{-1}}$ in umbral cores. The obtained velocities show a significant inverse dependence on the magnetic field strength.

We performed spectroscopic sunspot observations with LARS at an unprecedented spectral accuracy. The unique combination of a high-resolution spectrograph and a laser frequency comb for the absolute wavelength calibration, paired with a careful ephemeris calculation and the measurement of the laboratory wavelength, allowed us to determine accurate Doppler shifts of the \ion{Ti}{i}\,5713.9\,\AA\ line with an uncertainty of around $\mathrm{5\,m\,s^{-1}}$. In this way, we were able to the ascertain the exact amount of convective blueshift in a 3\arcsec-wide region of sunspot umbrae and verify the fundamental laws of magneto-convection. Compared to quiet Sun observations, we find a strong reduction of the convective blueshift by around $\mathrm{300\,m\,s^{-1}}$ for the observed umbrae. However, we detect that even the darkest parts of umbrae still possess convective blueshifts of a few tens of $\mathrm{m\,s^{-1}}$. A statistical analysis revealed a significant linear correlation between the convective blueshift and the magnetic field strength of sunspot umbrae. Thus, the absolute velocity decreases with increasing field strength from around $\mathrm{70\,m\,s^{-1}}$ at 2.2\,kG to around $\mathrm{20\,m\,s^{-1}}$ at 2.6\,kG. Following this trend, the convective blueshift in umbrae vanishes for field strengths of around 2.78\,kG. Above that value, we consider the magnetic field to be sufficiently dense to entirely suppress any sub-photospheric convection. We note that the stated values for the magnetic field strength base on HMI inversions of the \ion{Fe}{i}\,6173\,\AA\ line and might not be directly applicable to observations with other instruments or lines.

The fact that the darkest umbral regions exhibit a residual convective blueshift in the lower photosphere is an indication that the umbral heat flux is provided not only by radiation, but (to a small amount) also by convection. The composition of the umbral heat flux by radiative and convective energy transport is in line with magneto-hydrostatic models, for example by \citet{1965ApJ...141..548D}. Although we have not noticed umbral dots within the spectroscopically observed area, we cannot exclude their existence with seizes below the resolution limit of the telescope. The theoretical model of umbral gaps \citep{1979ApJ...234..333P} suggests that umbral dots are the observed hot tips of a dense pattern of field-free gaps just below the umbral surface. Within umbral gaps, the heat flux is carried by field-free convection. However, the gaps close near the continuum optical depth unity level of the magnetically dominated regions, which is why convective flows would be absent or very slow in the atmosphere above. The umbral gap model managed to bring the residual brightness of the umbra in line with zero convective blueshifts. The heat flux problem of the umbra appeared to be solved by the collective radiation from umbral dots. In a different approach, \citet{1989ApJ...342.1158M} theoretically estimated the amount of unresolved convection which would be necessary to provide the umbral heat flux. For the deep umbra, they expect a convective blueshift of around $\mathrm{-150\,m\,s^{-1}}$. 

On the basis of our resultant umbral velocities, we conclude that convective energy transport is also present in the umbra. We can only speculate that the heat flux is concentrated in unresolved umbral dots or gaps. The dependence of the blueshift on the magnetic field strength could be explained by the density of the umbral gap pattern. With increasing magnetic flux density, the gaps in the umbral core get less frequent which would then lead to a smaller amount of convective heat flux and therefore reduced convective blueshift. Based on our work, a new model of umbral cores might include the refinement of existing umbra models \citep[e.g.,][and references therein]{1992ASIC..375..103M,1997ASPC..118..122J} by a field-dependent magneto-convection reaching into the photosphere.

Our observations yield self-consistent and conclusive results on the convective shift in sunspot umbrae. We plan to perform further observations with LARS in the near future to increase the statistical sample. Thereby, we will alternate the instrument's field aperture between 3\arcsec\ and 1\arcsec. By reducing the statistical probability of umbral dots in our spectrally unresolved region, we aim to approach the zero convective blueshift of the umbral core. Since the laser frequency comb provides the absolute wavelength calibration and therefore guarantees identical instrumental conditions, existing and new measurements will be directly comparable. Our scheme is therefore only limited by the Sun itself, which is approaching its activity minimum with a smaller probability of sunspot occurrence. We will follow our approach of a spectroscopic analysis with the \ion{Ti}{i}\,5713.9\,\AA\ line. In comparison, we will also perform observation with the similarly suited and unblended \ion{Fe}{i}\,5576\,\AA\ line \citep[$\mathrm{g_{eff}=0}$,][]{2005A&A...439..687C}. With respect to the \ion{Fe}{i} line, the \ion{Ti}{i} line has the advantage that it gets substantially stronger in line depth in the cool umbra. This limits the impact of stray light from the much brighter quiet photosphere which plagues umbral observations \citep{1992SoPh..140..207M}. To reduce the amount of stray light to a minimum, we have restricted our investigation to sunspots with umbral diameters greater than 10\arcsec\ which are demonstrably less affected by scattered light \citep{1965smss.book.....Z,2007A&A...465..291M}. In our study, we have not addressed the impact of stray light on our line profiles. \citet{Doerr2015} estimates the fraction of light from the umbral surroundings to be around 5\%.

We agree to the velocity calibration of a sunspot region taking the dark umbra as reference. But the umbra may not be associated to be at rest with a zero velocity. Instead, the linear relation between the umbral velocity and the magnetic field strength provides the basis for an enhanced calibration. A possible application is the calibration of HMI Dopplergrams to an improved accuracy. Since HMI Dopplergrams and magnetic field inversions are steadily acquired and publicly available since 2010, the sample of sunspots is large. Due to the presence of p-mode oscillations, the umbral reference should base on a temporal average of around one hour. In a future study, we will present the results of the recalibration of sunspot Dopplergrams based on the magnetic field strength. We believe that this novel approach is of great interest for future studies of flow fields in active regions. 

\begin{acknowledgements} 
We thank all our colleagues at the Kiepenheuer Institute for Solar Physics, at Menlo Systems GmbH, and at the Max Planck Institute of Quantum Optics who worked on the development of the instrument. The LARS instrument and the Vacuum Tower Telescope at the Observatorio del Teide on Tenerife are operated by the Kiepenheuer Institute for Solar Physics Freiburg, which is a public law foundation of the State of Baden-W\"urttemberg. This work is part of a Post-doc project funded by the Deutsche Forschungsgemeinschaft (DFG, Ref.-No. Schm-1168/10). The initial astro-comb project for the VTT had been funded by the Leibniz-Gemeinschaft through the "Pakt f\"ur Forschung und Innovation". Finally, we want to thank Dr. Morten Franz for his fruitful comments on the manuscript.
\end{acknowledgements}
\bibliographystyle{aa} 
\bibliography{LARS}

\begin{thebibliography}{63}
\expandafter\ifx\csname natexlab\endcsname\relax\def\natexlab#1{#1}\fi

\bibitem[{{Acton}(1996)}]{Acton1996}
{Acton}, C.~H. 1996, Planetary and Space Science, 44, 65

\bibitem[{{Balthasar}(1984)}]{1984SoPh...93..219B}
{Balthasar}, H. 1984, \solphys, 93, 219

\bibitem[{{Balthasar} {et~al.}(1982){Balthasar}, {Thiele}, \&
  {Woehl}}]{Balthasar+etal1982}
{Balthasar}, H., {Thiele}, U., \& {Woehl}, H. 1982, \aap, 114, 357

\bibitem[{{Beckers}(1977)}]{Beckers1977}
{Beckers}, J.~M. 1977, \apj, 213, 900

\bibitem[{{Bellot Rubio} {et~al.}(2004){Bellot Rubio}, {Balthasar}, \&
  {Collados}}]{2004A&A...427..319B}
{Bellot Rubio}, L.~R., {Balthasar}, H., \& {Collados}, M. 2004, \aap, 427, 319

\bibitem[{{Bellot Rubio} {et~al.}(2008){Bellot Rubio}, {Tritschler}, \&
  {Mart{\'{\i}}nez Pillet}}]{2008ApJ...676..698B}
{Bellot Rubio}, L.~R., {Tritschler}, A., \& {Mart{\'{\i}}nez Pillet}, V. 2008,
  \apj, 676, 698

\bibitem[{{Biermann}(1941)}]{1941VAG....76..194B}
{Biermann}, L. 1941, Vierteljahresschrift der Astronomischen Gesellschaft, 76,
  194

\bibitem[{{Borrero} \& {Bellot Rubio}(2002)}]{2002A&A...385.1056B}
{Borrero}, J.~M. \& {Bellot Rubio}, L.~R. 2002, \aap, 385, 1056

\bibitem[{{Borrero} {et~al.}(2011){Borrero}, {Tomczyk}, {Kubo},
  {Socas-Navarro}, {Schou}, {Couvidat}, \& {Bogart}}]{2011SoPh..273..267B}
{Borrero}, J.~M., {Tomczyk}, S., {Kubo}, M., {et~al.} 2011, \solphys, 273, 267

\bibitem[{{Bruning}(1980)}]{1980A&A....81...50B}
{Bruning}, D.~H. 1980, \aap, 81, 50

\bibitem[{{Cabrera Solana} {et~al.}(2005){Cabrera Solana}, {Bellot Rubio}, \&
  {del Toro Iniesta}}]{2005A&A...439..687C}
{Cabrera Solana}, D., {Bellot Rubio}, L.~R., \& {del Toro Iniesta}, J.~C. 2005,
  \aap, 439, 687

\bibitem[{{de la Cruz Rodr{\'{\i}}guez} {et~al.}(2011){de la Cruz
  Rodr{\'{\i}}guez}, {Kiselman}, \& {Carlsson}}]{2011A&A...528A.113D}
{de la Cruz Rodr{\'{\i}}guez}, J., {Kiselman}, D., \& {Carlsson}, M. 2011,
  \aap, 528, A113

\bibitem[{{Deinzer}(1965)}]{1965ApJ...141..548D}
{Deinzer}, W. 1965, \apj, 141, 548

\bibitem[{{Doerr}(2015)}]{Doerr2015}
{Doerr}, H.-P. 2015, PhD thesis, University of Freiburg

\bibitem[{{Dravins} {et~al.}(1981){Dravins}, {Lindegren}, \&
  {Nordlund}}]{1981A&A....96..345D}
{Dravins}, D., {Lindegren}, L., \& {Nordlund}, A. 1981, \aap, 96, 345

\bibitem[{{Esteban Pozuelo} {et~al.}(2015){Esteban Pozuelo}, {Bellot Rubio}, \&
  {de la Cruz Rodr{\'{\i}}guez}}]{2015ApJ...803...93E}
{Esteban Pozuelo}, S., {Bellot Rubio}, L.~R., \& {de la Cruz Rodr{\'{\i}}guez},
  J. 2015, \apj, 803, 93

\bibitem[{{Forsberg}(1991)}]{1991PhyS...44..446F}
{Forsberg}, P. 1991, \physscr, 44, 446

\bibitem[{{Franz} \& {Schlichenmaier}(2009)}]{2009A&A...508.1453F}
{Franz}, M. \& {Schlichenmaier}, R. 2009, \aap, 508, 1453

\bibitem[{{Grossmann-Doerth} {et~al.}(1986){Grossmann-Doerth}, {Schmidt}, \&
  {Schroeter}}]{1986A&A...156..347G}
{Grossmann-Doerth}, U., {Schmidt}, W., \& {Schroeter}, E.~H. 1986, \aap, 156,
  347

\bibitem[{{Jahn}(1997)}]{1997ASPC..118..122J}
{Jahn}, K. 1997, in Astronomical Society of the Pacific Conference Series, Vol.
  118, 1st Advances in Solar Physics Euroconference. Advances in Physics of
  Sunspots, ed. B.~{Schmieder}, J.~C. {del Toro Iniesta}, \& M.~{Vazquez}, 122

\bibitem[{{Katsukawa} {et~al.}(2007){Katsukawa}, {Yokoyama}, {Berger},
  {Ichimoto}, {Kubo}, {Lites}, {Nagata}, {Shimizu}, {Shine}, {Suematsu},
  {Tarbell}, {Title}, \& {Tsuneta}}]{2007PASJ...59S.577K}
{Katsukawa}, Y., {Yokoyama}, T., {Berger}, T.~E., {et~al.} 2007, \pasj, 59,
  S577

\bibitem[{{Kleint} \& {Sainz Dalda}(2013)}]{2013ApJ...770...74K}
{Kleint}, L. \& {Sainz Dalda}, A. 2013, \apj, 770, 74

\bibitem[{{Koch}(1984)}]{1984SoPh...93...53K}
{Koch}, A. 1984, \solphys, 93, 53

\bibitem[{Kramida {et~al.}(2015)Kramida, {Yu.~Ralchenko}, Reader, \& {and NIST
  ASD Team}}]{NIST_ASD}
Kramida, A., {Yu.~Ralchenko}, Reader, J., \& {and NIST ASD Team}. 2015, {NIST
  Atomic Spectra Database (ver. 5.3), [Online]. Available:
  {\tt{http://physics.nist.gov/asd}} [2017, March 24]. National Institute of
  Standards and Technology, Gaithersburg, MD.}

\bibitem[{{Langangen} {et~al.}(2007){Langangen}, {Carlsson}, {Rouppe van der
  Voort}, \& {Stein}}]{2007ApJ...655..615L}
{Langangen}, {\O}., {Carlsson}, M., {Rouppe van der Voort}, L., \& {Stein},
  R.~F. 2007, \apj, 655, 615

\bibitem[{{Lites} {et~al.}(1991){Lites}, {Bida}, {Johannesson}, \&
  {Scharmer}}]{1991ApJ...373..683L}
{Lites}, B.~W., {Bida}, T.~A., {Johannesson}, A., \& {Scharmer}, G.~B. 1991,
  \apj, 373, 683

\bibitem[{{L{\"o}hner-B{\"o}ttcher}(2016)}]{2016PhDT........15L}
{L{\"o}hner-B{\"o}ttcher}, J. 2016, PhD thesis, Universit{\"a}t Freiburg im
  Breisgau

\bibitem[{{L{\"o}hner-B{\"o}ttcher} {et~al.}(2017){L{\"o}hner-B{\"o}ttcher},
  {Schmidt}, {Doerr}, {Kentischer}, {Steinmetz}, {Probst}, \&
  {Holzwarth}}]{2017A&A...607A..12L}
{L{\"o}hner-B{\"o}ttcher}, J., {Schmidt}, W., {Doerr}, H.-P., {et~al.} 2017,
  \aap, 607, A12

\bibitem[{{Maltby}(1992)}]{1992ASIC..375..103M}
{Maltby}, P. 1992, in NATO Advanced Science Institutes (ASI) Series C, Vol.
  375, NATO Advanced Science Institutes (ASI) Series C, ed. J.~H. {Thomas} \&
  N.~O. {Weiss}, 103--120

\bibitem[{{Martinez Pillet}(1992)}]{1992SoPh..140..207M}
{Martinez Pillet}, V. 1992, \solphys, 140, 207

\bibitem[{{Mart{\'{\i}}nez Pillet} {et~al.}(1997){Mart{\'{\i}}nez Pillet},
  {Lites}, \& {Skumanich}}]{1997ApJ...474..810M}
{Mart{\'{\i}}nez Pillet}, V., {Lites}, B.~W., \& {Skumanich}, A. 1997, \apj,
  474, 810

\bibitem[{{Mathew} {et~al.}(2007){Mathew}, {Mart{\'{\i}}nez Pillet}, {Solanki},
  \& {Krivova}}]{2007A&A...465..291M}
{Mathew}, S.~K., {Mart{\'{\i}}nez Pillet}, V., {Solanki}, S.~K., \& {Krivova},
  N.~A. 2007, \aap, 465, 291

\bibitem[{{Moreno-Insertis} \& {Spruit}(1989)}]{1989ApJ...342.1158M}
{Moreno-Insertis}, F. \& {Spruit}, H.~C. 1989, \apj, 342, 1158

\bibitem[{{Ortiz} {et~al.}(2010){Ortiz}, {Bellot Rubio}, \& {Rouppe van der
  Voort}}]{2010ApJ...713.1282O}
{Ortiz}, A., {Bellot Rubio}, L.~R., \& {Rouppe van der Voort}, L. 2010, \apj,
  713, 1282

\bibitem[{{Parker}(1979)}]{1979ApJ...234..333P}
{Parker}, E.~N. 1979, \apj, 234, 333

\bibitem[{{Pearson}(1896)}]{Pearson253}
{Pearson}, K. 1896, Philosophical Transactions of the Royal Society of London
  A: Mathematical, Physical and Engineering Sciences, 187, 253

\bibitem[{{Pesnell} {et~al.}(2012){Pesnell}, {Thompson}, \&
  {Chamberlin}}]{2012SoPh..275....3P}
{Pesnell}, W.~D., {Thompson}, B.~J., \& {Chamberlin}, P.~C. 2012, \solphys,
  275, 3

\bibitem[{{Rezaei} {et~al.}(2006){Rezaei}, {Schlichenmaier}, {Beck}, \& {Bellot
  Rubio}}]{2006A&A...454..975R}
{Rezaei}, R., {Schlichenmaier}, R., {Beck}, C., \& {Bellot Rubio}, L.~R. 2006,
  \aap, 454, 975

\bibitem[{{Riethm{\"u}ller} {et~al.}(2008){Riethm{\"u}ller}, {Solanki}, \&
  {Lagg}}]{2008ApJ...678L.157R}
{Riethm{\"u}ller}, T.~L., {Solanki}, S.~K., \& {Lagg}, A. 2008, \apjl, 678,
  L157

\bibitem[{{Riethm{\"u}ller} {et~al.}(2013){Riethm{\"u}ller}, {Solanki}, {van
  Noort}, \& {Tiwari}}]{2013A&A...554A..53R}
{Riethm{\"u}ller}, T.~L., {Solanki}, S.~K., {van Noort}, M., \& {Tiwari}, S.~K.
  2013, \aap, 554, A53

\bibitem[{{Rimmele}(1994)}]{1994A&A...290..972R}
{Rimmele}, T.~R. 1994, \aap, 290, 972

\bibitem[{{Rimmele}(2004)}]{2004ApJ...604..906R}
{Rimmele}, T.~R. 2004, \apj, 604, 906

\bibitem[{{S{\'a}nchez Almeida}(2005)}]{2005ApJ...622.1292S}
{S{\'a}nchez Almeida}, J. 2005, \apj, 622, 1292

\bibitem[{{Schmidt} \& {Balthasar}(1994)}]{Schmidt+Balthasar1994}
{Schmidt}, W. \& {Balthasar}, H. 1994, \aap, 283, 241

\bibitem[{{Schou} {et~al.}(2012){Schou}, {Scherrer}, {Bush}, {Wachter},
  {Couvidat}, {Rabello-Soares}, {Bogart}, {Hoeksema}, {Liu}, {Duvall}, {Akin},
  {Allard}, {Miles}, {Rairden}, {Shine}, {Tarbell}, {Title}, {Wolfson},
  {Elmore}, {Norton}, \& {Tomczyk}}]{2012SoPh..275..229S}
{Schou}, J., {Scherrer}, P.~H., {Bush}, R.~I., {et~al.} 2012, \solphys, 275,
  229

\bibitem[{{Schr\"{o}ter} {et~al.}(1985){Schr\"{o}ter}, {Soltau}, \&
  {Wiehr}}]{1985VA.....28..519S}
{Schr\"{o}ter}, E.~H., {Soltau}, D., \& {Wiehr}, E. 1985, Vistas in Astronomy,
  28, 519

\bibitem[{{Sch{\"u}ssler} \& {V{\"o}gler}(2006)}]{2006ApJ...641L..73S}
{Sch{\"u}ssler}, M. \& {V{\"o}gler}, A. 2006, \apjl, 641, L73

\bibitem[{Shapiro \& Wilk(1965)}]{normality_test}
Shapiro, S.~S. \& Wilk, M.~B. 1965, Biometrika, 52, 591

\bibitem[{{Shine} {et~al.}(1994){Shine}, {Title}, {Tarbell}, {Smith}, {Frank},
  \& {Scharmer}}]{1994ApJ...430..413S}
{Shine}, R.~A., {Title}, A.~M., {Tarbell}, T.~D., {et~al.} 1994, \apj, 430, 413

\bibitem[{{Sigwarth} {et~al.}(1999){Sigwarth}, {Balasubramaniam},
  {Kn{\"o}lker}, \& {Schmidt}}]{1999A&A...349..941S}
{Sigwarth}, M., {Balasubramaniam}, K.~S., {Kn{\"o}lker}, M., \& {Schmidt}, W.
  1999, \aap, 349, 941

\bibitem[{{Snodgrass} \& {Ulrich}(1990)}]{1990ApJ...351..309S}
{Snodgrass}, H.~B. \& {Ulrich}, R.~K. 1990, \apj, 351, 309

\bibitem[{{Sobotka} \& {Jur{\v c}{\'a}k}(2009)}]{2009ApJ...694.1080S}
{Sobotka}, M. \& {Jur{\v c}{\'a}k}, J. 2009, \apj, 694, 1080

\bibitem[{{Socas-Navarro} {et~al.}(2004){Socas-Navarro}, {Mart{\'{\i}}nez
  Pillet}, {Sobotka}, \& {V{\'a}zquez}}]{2004ApJ...614..448S}
{Socas-Navarro}, H., {Mart{\'{\i}}nez Pillet}, V., {Sobotka}, M., \&
  {V{\'a}zquez}, M. 2004, \apj, 614, 448

\bibitem[{{Solanki}(1986)}]{1986A&A...168..311S}
{Solanki}, S.~K. 1986, \aap, 168, 311

\bibitem[{{Stanchfield} {et~al.}(1997){Stanchfield}, {Thomas}, \&
  {Lites}}]{1997ApJ...477..485S}
{Stanchfield}, II, D.~C.~H., {Thomas}, J.~H., \& {Lites}, B.~W. 1997, \apj,
  477, 485

\bibitem[{{Steinmetz} {et~al.}(2008){Steinmetz}, {Wilken}, {Araujo-Hauck},
  {Holzwarth}, {H{\"a}nsch}, {Pasquini}, {Manescau}, {D'Odorico}, {Murphy},
  {Kentischer}, {Schmidt}, \& {Udem}}]{Steinmetz+etal2008}
{Steinmetz}, T., {Wilken}, T., {Araujo-Hauck}, C., {et~al.} 2008, Science, 321,
  1335

\bibitem[{{Thomas} {et~al.}(1984){Thomas}, {Cram}, \&
  {Nye}}]{1984ApJ...285..368T}
{Thomas}, J.~H., {Cram}, L.~E., \& {Nye}, A.~H. 1984, \apj, 285, 368

\bibitem[{{Tritschler} {et~al.}(2004){Tritschler}, {Schlichenmaier}, {Bellot
  Rubio}, {KAOS Team}, {Berkefeld}, \& {Schelenz}}]{2004A&A...415..717T}
{Tritschler}, A., {Schlichenmaier}, R., {Bellot Rubio}, L.~R., {et~al.} 2004,
  \aap, 415, 717

\bibitem[{{von der L\"{u}he} {et~al.}(2003){von der L\"{u}he}, {Soltau},
  {Berkefeld}, \& {Schelenz}}]{2003SPIE.4853..187V}
{von der L\"{u}he}, O., {Soltau}, D., {Berkefeld}, T., \& {Schelenz}, T. 2003,
  in \procspie, Vol. 4853, Innovative Telescopes and Instrumentation for Solar
  Astrophysics, ed. S.~L. {Keil} \& S.~V. {Avakyan}, 187--193

\bibitem[{{Watanabe} {et~al.}(2012){Watanabe}, {Bellot Rubio}, {de la Cruz
  Rodr{\'{\i}}guez}, \& {Rouppe van der Voort}}]{2012ApJ...757...49W}
{Watanabe}, H., {Bellot Rubio}, L.~R., {de la Cruz Rodr{\'{\i}}guez}, J., \&
  {Rouppe van der Voort}, L. 2012, \apj, 757, 49

\bibitem[{{Watanabe} {et~al.}(2009){Watanabe}, {Kitai}, \&
  {Ichimoto}}]{2009ApJ...702.1048W}
{Watanabe}, H., {Kitai}, R., \& {Ichimoto}, K. 2009, \apj, 702, 1048

\bibitem[{{W{\"o}ger} \& {von der L{\"u}he}(2008)}]{2008SPIE.7019E..1EW}
{W{\"o}ger}, F. \& {von der L{\"u}he}, II, O. 2008, in Society of Photo-Optical
  Instrumentation Engineers (SPIE) Conference Series, Vol. 7019, Society of
  Photo-Optical Instrumentation Engineers (SPIE) Conference Series, 1

\bibitem[{{Zwaan}(1965)}]{1965smss.book.....Z}
{Zwaan}, C. 1965, {Sunspot models; a study of sunspot spectra}

\end{thebibliography}

\clearpage
\begin{appendix}
\section{Additional figures}

\begin{figure}[htpb]
\includegraphics[trim=1.7cm 4.9cm 9.75cm 3.0cm,clip,width=0.95\columnwidth]{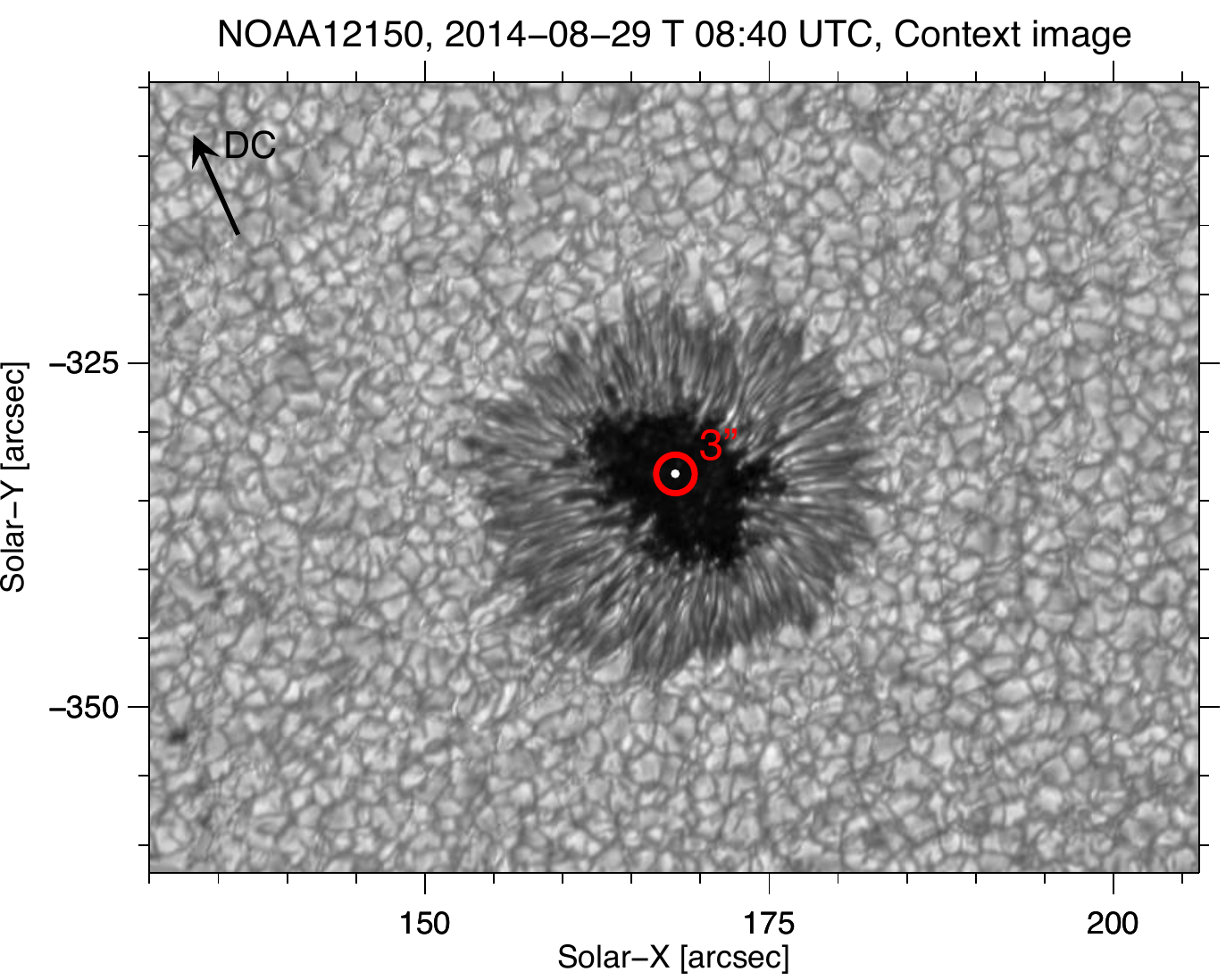}
\caption{Sunspot of NOAA\,12150 observed with the LARS context imager on August 29 2014 at 8:40\,UTC in the spectral G-band at 4307\,\AA. The black arrow is pointing in the direction of the solar disk center. The spectroscopically analyzed 3\arcsec-wide region which is highlighted by a red circle was centered to the darkest umbra at a heliographic position of $10.4^{\circ}$\,W and $13.4^{\circ}$\,S. For more information see Obs.\,1 in Table\,\ref{table_observations}.}
\label{fig_results_Ti5714_spotA1}
\end{figure}

\begin{figure}[htpb]
\includegraphics[trim=1.6cm 4.6cm 9.7cm 3.0cm,clip,width=0.95\columnwidth]{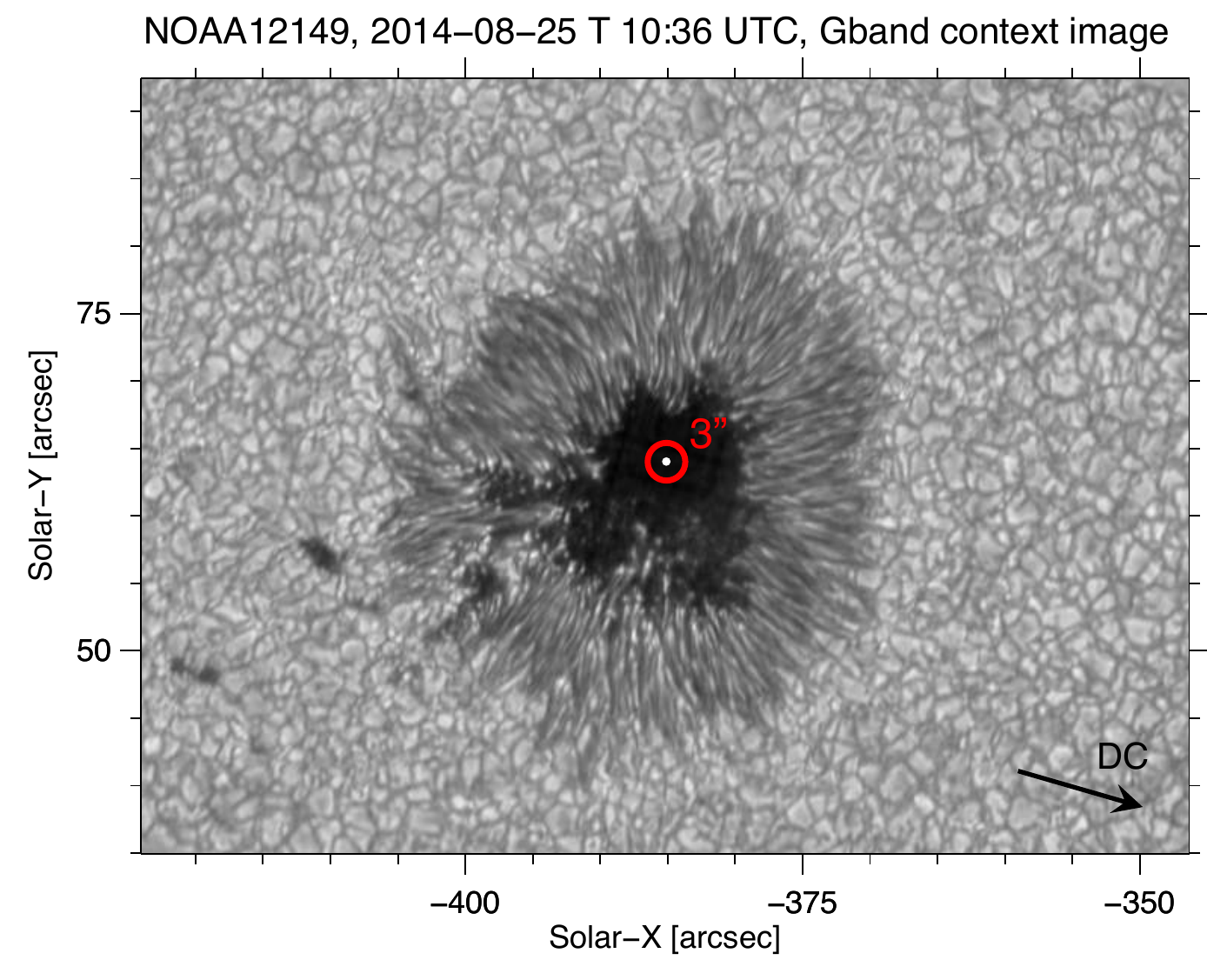}
\caption{Sunspot of NOAA\,12149 observed with the LARS context imager on August 25 2014 at 10:36\,UTC in the spectral G-band at 4307\,\AA. The black arrow is pointing in the direction of the solar disk center. The spectroscopically analyzed 3\arcsec-wide region which is highlighted by a red circle was centered to the darkest umbra at a heliographic position of $24.2^{\circ}$\,E and $10.3^{\circ}$\,N. For more information see Obs.\,5 in Table\,\ref{table_observations}.}
\label{fig_results_Ti5714_spotA2}
\end{figure}

\begin{figure}[htpb]
\includegraphics[trim=2cm 3.65cm 10cm 2.3cm,clip,width=0.85\columnwidth]{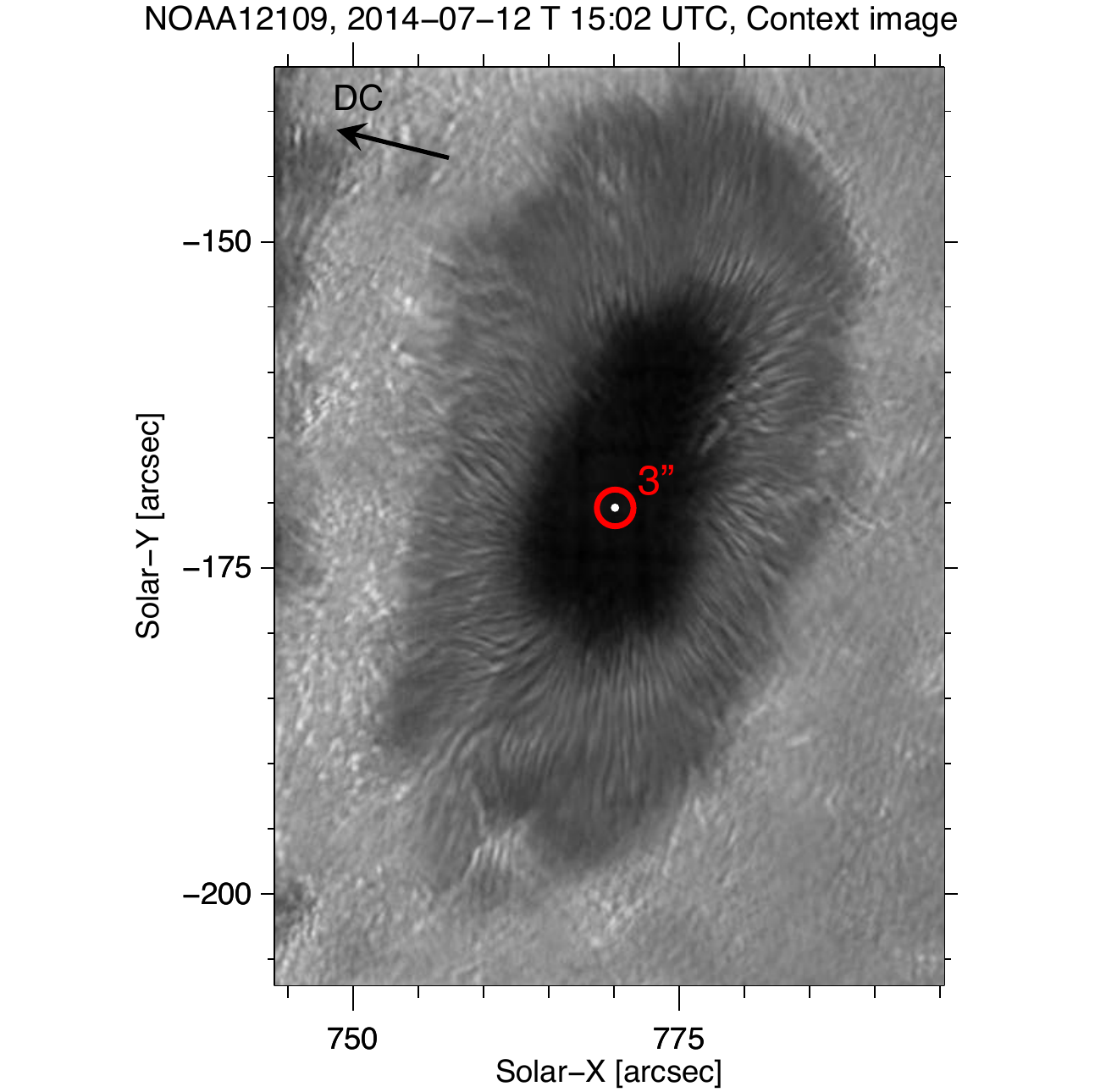}
\caption{Sunspot of NOAA\,12109 observed on July 12 2014 at 15:02\,UTC in the spectral G-band at 4307\,\AA. The black arrow is pointing in the direction of the solar disk center. The spectroscopically analyzed 3\arcsec-wide region (red circle) was centered at a heliographic position of $55.3^{\circ}$\,W and $8.1^{\circ}$\,S. For more information see Obs.\,13 in Table\,\ref{table_observations}.}
\label{fig_results_Ti5714_spotA3}
\end{figure}

\begin{figure}[htpb]
\includegraphics[width=0.96\columnwidth]{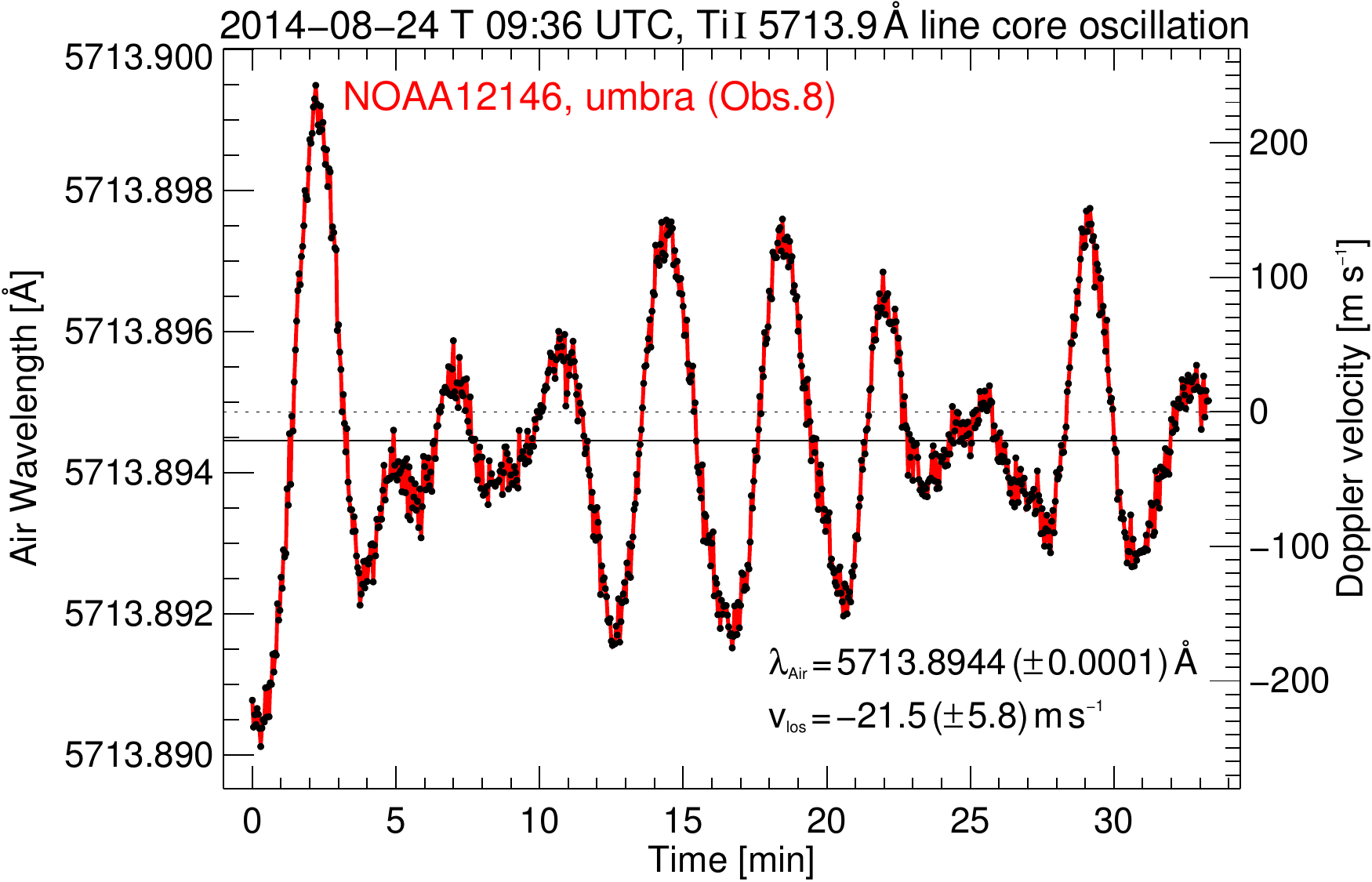}
\caption{Temporal variation of the umbral Doppler shifts of an observation sequence (see Obs.\,8 in Table\,\ref{table_observations}) The points represent the position of the \ion{Ti}{I} line center in absolute air wavelength (left y-axis) and respective Doppler velocities in $\mathrm{m\,s^{-1}}$ (right y-axis). The temporal average at 5713.8944\,\AA\ is marked by a black solid line and corresponds to a blueshift of $\mathrm{-21.5\,m\,s^{-1}}$. The error of the mean is given in brackets.}
\label{fig_results_Ti5714_oscillation}
\end{figure}

\begin{figure}[htpb]
\includegraphics[width=0.96\columnwidth]{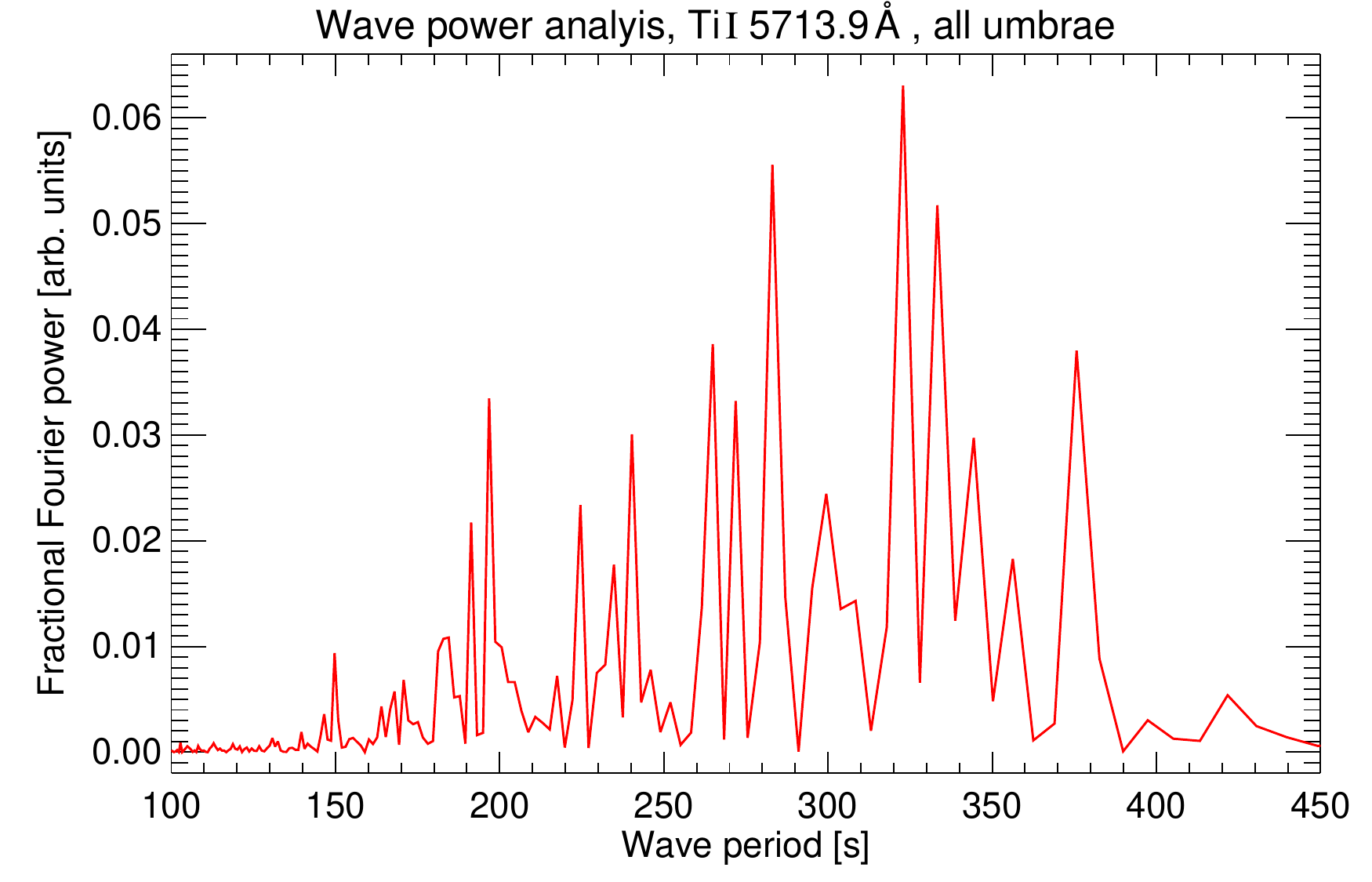}
\caption{Global power spectrum of umbral waves from all datasets. The fractional Fourier power is plotted as a function of the wave period.}
\label{fig_results_frequency_analysis}
\end{figure}

\end{appendix}
\end{document}